\documentclass{article}
\usepackage{natbib,emulateapj,onecolfloat}

\newcommand{\rbin}{r_{12}}
\newcommand{\NB}{{\sf NBODY6++}\ }
\newcommand{\GDT}{{\sf GADGET}\ }

\newcommand{\mh}{M_{\bullet}}

\newcommand{\ms}{\mh-\sigma}
\newcommand{\mbin}{M_{12}}
\newcommand{\msun}{M_{\sun}}

\newcommand{\beq}{\begin{equation}}
\newcommand{\eeq}{\end{equation}}
\newcommand{\erf}{\textrm{erf}}
\newcommand{\erfp}{\textrm{erf}\;'}

\begin{document}
\twocolumn
[
\title{Formation of Galactic Nuclei}
\author{Milo\v s Milosavljevi\' c and David Merritt}
\affil{Department of Physics and Astronomy, Rutgers University}
\authoremail{milos@physics.rutgers.edu}
\authoremail{merritt@physics.rutgers.edu}

\begin{abstract}
\noindent
We investigate a model in which galactic nuclei form via the 
coalescence of pre-existing stellar systems containing supermassive 
black holes.
Merger simulations are carried out using $N$-body algorithms that can
follow the formation and decay of a black-hole binary and its effect on the 
surrounding stars down to sub-parsec scales. 
Our initial stellar systems have steep central density cusps similar to those
in low-luminosity elliptical galaxies.
Immediately following the merger, the density profile of the remnant
is homologous with the initial density profile
and the steep nuclear cusp is preserved.
However the formation of a black-hole binary transfers energy to the stars
and lowers the central density;
continued decay of the binary creates a $\rho\sim r^{-1}$ density
cusp similar to those observed in bright elliptical galaxies,
with a break radius that extends well beyond the sphere of
gravitational influence of the black holes.
Our simulations are the first to successfully produce shallow
power-law cusps from mergers of galaxies with steep cusps, 
and our results support a picture in which the observed dependence of 
nuclear cusp slope on galaxy luminosity is a consequence of 
galaxy interactions.
We discuss the implications of our results for the survivability
of dark-matter cusps.

We follow the decay of the black hole binary over a factor of
$\sim 20$ in separation after formation of a hard binary,
considerably farther than in previous simulations.
We see almost no dependence of the binary's decay rate
on number of particles in the simulation, contrary
to earlier studies in which a lower initial density of stars
led to a more rapid depletion of the binary's loss cone.
We nevertheless argue that the decay of a black hole binary
in a real galaxy would be expected to stall at 
separations of $0.01-1$ pc unless some additional mechanism 
is able to extract energy from the binary.

\end{abstract}
]
\section{Introduction}
\label{sec_intro}

Galactic nuclei\footnote[1]{We use the term ``nucleus'' 
to refer generically to the central parts of galaxies.
The term is sometimes used more restrictively to refer to 
pointlike nuclei, e.g. \citet{kom93}.}
are regions of high stellar density at the centers of galaxies.
Early studies of the evolution of galactic nuclei 
\citep{sps66, sps67, col67, san70}
emphasized stellar encounters and collisions as the dominant 
physical processes.
In these models, 
the density of a compact stellar system gradually increases as energetic 
stars are scattered into elongated orbits via two-body encounters.
The increase in density leads to a higher rate of 
physical collisions between stars, 
liberating gas that falls to the center of the system and 
condenses into new stars which undergo further collisions.
\citet{ber78} argued that the evolution of a dense
nucleus would lead inevitably to the formation of a massive black hole (BH) 
at the center, either by runaway stellar mergers or by
creation of a massive gas cloud which collapses.
Subsequent studies \citep{dus83,qus87,qus89,ddc87a,ddc87b,mcd91}
have included ``seed'' BHs which grow via accretion of stars
or gas liberated by stellar collisions.

The fundamental time scale in these models is the relaxation 
time determined by the stars, or
\beq
t_r \approx 0.34{\sigma_*^3\over G^2m_*\rho\ln\Lambda}
\label{eq_trelax}
\eeq
\citep{sph71},
where $\sigma_*$ is the stellar 1D velocity dispersion,
$m_*$ and $\rho$ are the stellar mass and mass density,
and $\ln\Lambda$ is the Coulomb logarithm.
Observations before about 1990 lacked the resolution to determine
whether $t_r$ was shorter than a Hubble time on scales smaller
than $\sim 1$ pc in galactic nuclei.
We now know that stellar densities increase approximately as
power laws in galactic nuclei, $\rho\sim r^{-\gamma}$, 
down to the smallest radii that can be resolved,
or roughly $10^{-1}$ pc in the nearest galaxies.
For instance, the Local Group galaxies M31, M32 and the Milky Way
all have nuclear density cusps with $\gamma=1.5\pm 0.5$ \citep{lau98}.
Furthermore the evidence for supermassive BHs is compelling in
these galaxies.
Within the BH's sphere of influence $r_{gr}$, where
\beq
r_{gr} \equiv {G\mh\over\sigma_*^2} \approx 10.8\: \mathrm{pc} 
\left({\mh\over 10^8\msun}\right) 
\left({\sigma_*\over 200\ \mathrm{ km}\ \mathrm{ s}^{-1}}\right)^{-2},
\eeq
the stellar velocity dispersion rises as $\sim r^{-1/2}$.
Equation (\ref{eq_trelax}) then implies $t_r\propto r^{\gamma-3/2}
\approx r^0$,
nearly independent of radius.
The relaxation time based on observations at $\sim 0.1''$ exceeds
$10^{11}$ yr in almost all galaxies \citep{fab97};
this angular size corresponds roughly to a radius $r_{gr}$ for
nearby galaxies \citep{mef01b},
from which it follows that $t_r$ is likely to exceed a Hubble
time at smaller radii as well.
(The pointlike nucleus of M33 is probably an exception; 
see \citet{kom93}.)

Physical collisions between stars occur on a timescale
that is longer than $t_r$ by roughly a factor 
$(\ln\Lambda)\Theta^2/(1+\Theta)\approx 10$ where 
$\Theta$, the ``Safronov number'' \citep{saf60},
is of order a few for stars in a galactic nucleus.
Thus neither elastic nor inelastic gravitational encounters 
are likely to be of dominant importance in determining the structure 
of nuclei containing supermassive BHs.

Nevertheless the properties of galactic nuclei do vary in
systematic ways with the properties of their host galaxies
\citep{kor85,lau85,fab97} and one would like to understand this.
Recent discussions of the formation and evolution of
galactic nuclei have begun from the assumption that supermassive
BHs were created during the quasar epoch and have been present ever since
with roughly their current masses.
Another element missing from the earlier studies was galactic mergers.
Mergers are complex phenomena, but an almost certain consequence of
a merger is the infall of the progenitor galaxy's BHs into the
nucleus of the merged system \citep{bbr80}.
An infalling BH would be expected to carry with it a mass in stars
of order its own mass, and decay of the BHs' orbits would inject a 
substantial amount of energy into the stars, 
enough to determine the structure of the remnant
nucleus out to a radius of several times $r_{gr}$
\citep{ebi91}.
In this picture, the structure and kinematics of galactic nuclei
are fossil relics of the merger histories of galaxies and of
the interaction between stars and supermassive binary BHs.

The present paper is a numerical study of this formation model.
We simulate the merger of two galaxies, each of which contains a central
point mass representing a supermassive BH.
Our study is unique in that it follows the details of the merger from its
earliest stages, when the two galaxies are distinct, to its late stages,
when the BHs have formed a hard binary and the binary has decayed via
energy exchange with stars to a separation much less than one parsec.
No existing $N$-body code can efficiently follow the evolution over
such a wide range of scales; 
hence we break the calculation into two parts, before and after
formation of the BH binary, and use different algorithms for each
(\S2).

We also include for the first time initial models which
are self-consistent realizations of galaxies with steep central density 
cusps, $\rho\sim r^{-2}$.
This choice is motivated by a number of lines of argument 
which suggest that the density of stars around a supermassive BH
should be a steep power law.
Random gravitational encounters between stars lead, over two-body
relaxation times, to density profiles of the form $\rho\sim r^{-2.23}$ 
in the absence of a black hole \citep{coh80}
and $\rho\sim r^{-1.75}$ in the presence of a BH \citep{baw76}.
It was argued above that relaxation processes are probably of
secondary importance in most nuclei, but even the slow growth of a BH 
in a pre-existing, collisionless nucleus produces a density
profile of the form $\rho\sim r^{-\gamma}$, $r\lesssim r_{gr}$ 
with $1.5\lesssim\gamma\lesssim 2.5$ \citep{pee72,you80,qhs95,vdm99}.
Low-luminosity elliptical galaxies and the bulges of spiral
galaxies are observed to have steep cusps,
$1.5\lesssim\gamma\lesssim 2.5$ \citep{fer94,lau95};
these galaxies are the least likely
to have been strongly affected by mergers and hence their density
profiles may reflect the structure of all nuclei at early times.
Finally, hierarchical growth of structure in the universe
generically produces systems with steep central density cusps,
$\rho\sim r^{-1.5}$ \citep{duc91,nfw96,moo98}, 
although simulations do not yet have sufficient resolution to make 
predictions on parsec or sub-parsec scales \citep{moo01}.

A major success of our study is the demonstration 
(\S\ref{sec_merge},\ref{sec_hard}) that
the merger of two galaxies with steep, power-law density cusps can produce
a galaxy with a shallow power-law cusp.
Shallow cusps (also called ``cuspy cores'') are observed in the brightest 
elliptical galaxies \citep{mef96,geb96},
and, 
while their origin has tentatively been associated with
the binary BH model \citep{ebi91,fab97},
no previous simulation had the resolution necessary to follow
the coalescence of initially steep cusps.
Our results support a model in which the observed dependence of 
nuclear cusp slope on galaxy luminosity is a consequence of 
galaxy interactions (\S\ref{sec_discuss}).

We also discuss in detail (\S\ref{sec_hard}) the decay of the BH binary;
we follow that decay over a factor $\sim 20$ in semimajor axis
after formation of a bound pair, considerably farther than
in earlier simulations.
An important question is the dependence of the binary hardening 
rate on $N$, the number of particles in the simulation.
Earlier studies had noted a decreasing decay rate with
increasing $N$, implying that the decay in real galaxies, 
where $N$ is very large, might be slow.
We do not observe an appreciable $N$ dependence in our simulations;
we discuss the likely reasons for this in \S\ref{sec_hard}, but argue
nevertheless that the decay would probably stall in real galaxies,
at separations of $0.01-1$ pc, unless some additional
mechanism is effective at extracting energy from the binary.
We are therefore led to predict (\S\ref{sec_discuss}) 
that some galaxies contain uncoalesced BH binaries at the current epoch.

We also present the morphological and kinematical structure of
the merged galaxy on sub-parsec scales (\S\ref{sec_kin}) 
and discuss some observational
signatures associated with our formation picture.

\section{Method}
\label{SectionMethod}

In this section we describe the initial conditions and algorithms used 
in our simulations and compare them with those of other authors.

\begin{deluxetable}{lcr}
\tablecolumns{3} 
\tablewidth{17pc}
\tablehead{\colhead{Parameter} & 
           \colhead{Symbol} & 
	   \colhead{Value}}
\tablecaption{Model Parameters\label{tab_model}}
\startdata
Mass of Galaxy & $M$ & $1$ \\ 
Mass of Black Hole & $\mh$ & $0.01$  \\ 
Half-Mass Radius & $r_0$ & $1$ \\ 
Total Energy & $E$ & $-0.25$ \\
Number of Stars & $N$ & $131,072$ \\
Stellar Mass & $m_*$ & $7.63\times10^{-6}$ \\
\enddata
\end{deluxetable}

Initial conditions consisted of twin, spherical stellar systems 
following the density law
\begin{equation}
\label{eq_jaffe}
\rho(r)=\frac{M}{4\pi r_0^3} \left(\frac{r}{r_0}\right)^{-2}
\left(1+\frac{r}{r_0}\right)^{-2}
\end{equation}
\citep{jaf83,deh93}, where $r_0$ is the half-mass radius and
$M$ the total mass.
To each of the models was added a central point of mass 
$\mh=0.01 M$ representing the supermassive BH.
The ratio $\mh/M$ in our models is somewhat greater than the mean
ratio of BH mass to galaxy mass in observed galaxies,
$\sim 1.2\times 10^{-3}$ \citep{mef01b};
however the radius of influence of our BHs in our merged galaxies, 
$r_{gr}=G\mh/\sigma_*^2\approx0.02$, is much smaller than $r_0$ which 
allows us to ignore the large-scale stellar distribution when scaling our
models to the nuclei of real galaxies; see \S\ref{sec_discuss}.
Velocities of the stars were generated from the unique, isotropic 
distribution function \citep{tre94} that reproduces the density law 
(\ref{eq_jaffe}) in the combined potential of the stars and the
central point mass.
Thus our models are initially in a state
of detailed equilibrium.
The values chosen for the model parameters are listed in Table 
\ref{tab_model}; Newton's constant is set to 
unity.\footnote{These parameters agree with the \citet{hem86}
``standard'' units.}  
The galaxies were set at time $t=0$ in an elliptic mutual orbit of 
semimajor axis $a_G=2$ and velocity $v_G=0.1425$ equal to half of the 
circular orbit velocity.  
(The index ``$G$'' labels the binary composed of two galaxies as distinct 
from a binary composed of two BHs.)
Since the relative orbit rapidly circularizes, we do not
expect our results to be strongly affected by the choice of
orbital initial conditions.

Our goal was to follow the evolution of this binary system from
its earliest stages, when the galaxies were distinct, through the
formation of a bound BH pair, until the gradual exchange 
of energy between BHs and stars had caused the BH binary to shrink
to sub-parsec separations.
No single $N$-body code currently in existence can deal efficiently
and accurately with evolution over such a broad range of scales;
particularly demanding is the treatment of BH-BH and BH-star
interactions, which require an algorithm that can accurately integrate
the equations of motion of point masses without softening.
The closest approximation to such a code is {\sf SCFBDY} developed by 
G. Quinlan at Rutgers University and used in two published studies
\citep{quh97,meq98}.
{\sf SCFBDY} combined elements of Aarseth's {\sf NBODY}
series of codes, including regularization of the BH-BH interaction,
with a low-resolution, mean-field potential solver for computing
the force field due to the stars.  
However {\sf SCFBDY} is only
suited to systems with a single density center and a high
degree of symmetry, ruling out its application to mergers.
{\sf SCFBDY} also uses softened gravity for computing the BH-star 
interactions, an approximation that affects the accuracy of the
critical interactions leading to the decay of the BH binary.
This code is also not available in a form that runs on parallel
architectures, limiting the number of particles that can be used.

We chose to break the problem into two parts, using a tree code
for the early stages of the merger (roughly until the formation 
of a BH binary), and a high-precision, direct-summation code for 
the later stages.
Ideally, one would use a high-precision code right from the 
start, to handle the steep force gradients produced
by the BHs and the other stars in the cusps.
However we were willing to accept some inaccuracy in the integrations
during the early stages of the merger if in so doing
we could treat a larger $N$; our only prerequisite was that
motions of stars on scales larger than the separation of the hard
BH binary should be faithfully tracked during the
early stages of the merger.
This reasoning motivated our choice of the recently-released
tree code \GDT \citep{syw01}
as the integrator for the early stages of the merger.
Features of \GDT that are relevant here include domain 
decomposition of the particle data set, mapping of particles onto 
the classic octal tree structure that respects hierarchical clustering 
of particles, quadrupolar expansion of force moments for spatially 
separated nodes, and individual and adaptive time steps for all particles.  
Force integration is controlled through the parameters $(h,\eta)$ that 
denote, respectively, the gravitational softening length and the time 
step accuracy factor, $\Delta t=\eta /|{\bf a}|$.  
Special care was taken to identify parameter values leading to optimum 
accuracy and efficiency on the Rutgers Sun HPC-10000 and the SDSC Cray T3E 
systems where the \GDT runs were produced.  
The softening length was chosen to be $h=0.001$, smaller than both the 
BH gravitational radius $r_{gr}\equiv G\mbin/\sigma_*^2\approx 0.01$ 
and the separation corresponding to a hard binary,
 $r_h\equiv G\mbin/8\sigma_*^2 \approx 0.0025$.
We monitored the density profiles and Lagrangian radii of the stellar cusps 
as diagnostics sensitive to corruption of the bulk stellar distribution 
due to unacceptable levels of softening.  
Runs entering final selection exhibited no cumulative distortions on 
scales larger than $\sim h$.
The total particle number used in the production run was 
$N=2^{18}=262,144$.  

\begin{deluxetable}{lccrr}
\tablecolumns{5} 
\tablewidth{30pc}
\tablehead{\colhead{Label} & 
	   \colhead{Reduction in Particle Number} & 
	   \colhead{Truncation in Energy} & 
	   \colhead{$N$} & 
	   \colhead{$\mh/m_*$}}
\tablecaption{\NB Initial Conditions\label{tab_runs}}
\startdata
\GDT & $1\times$ & $1\times$ & $262,144$ & $1,311$ \\
{\sf A2} & $2\times$ & $4\times$ & $32,768$ & $655$ \\
{\sf A4} & $4\times$ & $4\times$ & $16,384$ & $328$ \\
{\sf A8} & $8\times$ & $4\times$ & $8,192$ & $164$ \\
{\sf B2} & $2\times$ & $4\times$ & $32,768$ & $1,311$ \\
{\sf B4} & $4\times$ & $4\times$ & $16,384$ & $655$ \\
{\sf B8} & $8\times$ & $4\times$ & $8,192$ & $328$
\enddata
\end{deluxetable}

The late stages of the evolution were integrated using the direct-summation
code \NB \citep{spb99}.
Conceived for the study of relaxation phenomena in globular clusters,
\NB and its serial progenitor {\sf NBODY6} are the last and most
complex codes in the {\sf NBODYx} series of \citet{aar99},
employing the fourth-order Hermite scheme \citep{maa92}
as their primary integrator.
The codes were written to facilitate the exact integration
(no softening) of a large number of bodies with approximately
equal masses.
\NB gives particles adaptive block-individual time steps
$\Delta t_n\propto 2^{-n}$ that are short for particles in
dense regions and as much as $10^2$ times longer for particles at 
the outskirts of the system. 
The \citet{ahc73} scheme is used to select a subset of
neighbors whose forces on the test particle are
extrapolated at higher time resolution than those of the non-neighbors.
Near encounters are treated using the two-body KS regularization
scheme \citep{kus65} and its generalizations to
systems with a few bodies, including the triple, quad and chain regularizations.
Chain regularization is a systematic procedure for serializing the
pairwise KS regularization in a group of not more than
six bodies in close approach.
\NB was developed to run on low-latency distributed memory
architectures such as the Cray T3E.
There is no domain decomposition:
every processing node contains an identical copy of the whole
dataset.
Only the do-loops are broken into parallel sections;
after every force calculation, an all-to-all broadcast scheme
updates the particle sets.
The treatment of binaries (regularization etc.) is not parallel.
In the simulations conducted on the 272-node Cray T3E at the SDSC
and the 64-node Sun HPC-10000 at Rutgers,
\NB scaled well with the number of processors when the spread
in time steps was moderate.
When a few particles had much shorter time steps than others,
the scaling was poor and serial integration on 666 MHz Alpha chips
was found to be preferable.

One goal of our study was to identify any $N$-dependent features
of the evolution, since real galaxies have much larger numbers
of stars than accessible to direct simulation.
To isolate the effects of varying $N$ from other dependence
(and also because \NB can not deal with particle numbers
as large as $10^5$ on currently available machines),
we drew various random particle sets from the $N=2^{18}$
\GDT integration and used these as initial conditions for the
\NB runs.
Half of the particles were iteratively removed and the
masses of the remaining particles (except for the BHs)
were doubled.
We chose the time $t_0=10.6$ to select our reduced data sets;
the separation of the BHs at this time is $0.072$, 
substantially greater than their separation ($\sim 0.0025$) at the
time $t_h$ when a hard binary forms.

In addition, to increase the effective resolution near the center, 
we sorted the data set by energy $E_i = m_*[v_i^2/2+U({\bf r}_i)]$ 
and removed the upper $3/4$ of all stars.  
While the new data sets were statistically distinct from the original, 
we expect to find unchanged dynamics in the cusps where the fractional 
perturbation from the removed stars is negligible.  
Combination of these two techniques led to a set of initial conditions 
for \NB labeled according to the formula ``{\sf An}'' 
where $n$ is the fractional reduction (Table \ref{tab_runs}).

\begin{figure*}[t]
\epsscale{1.8}
\plotone{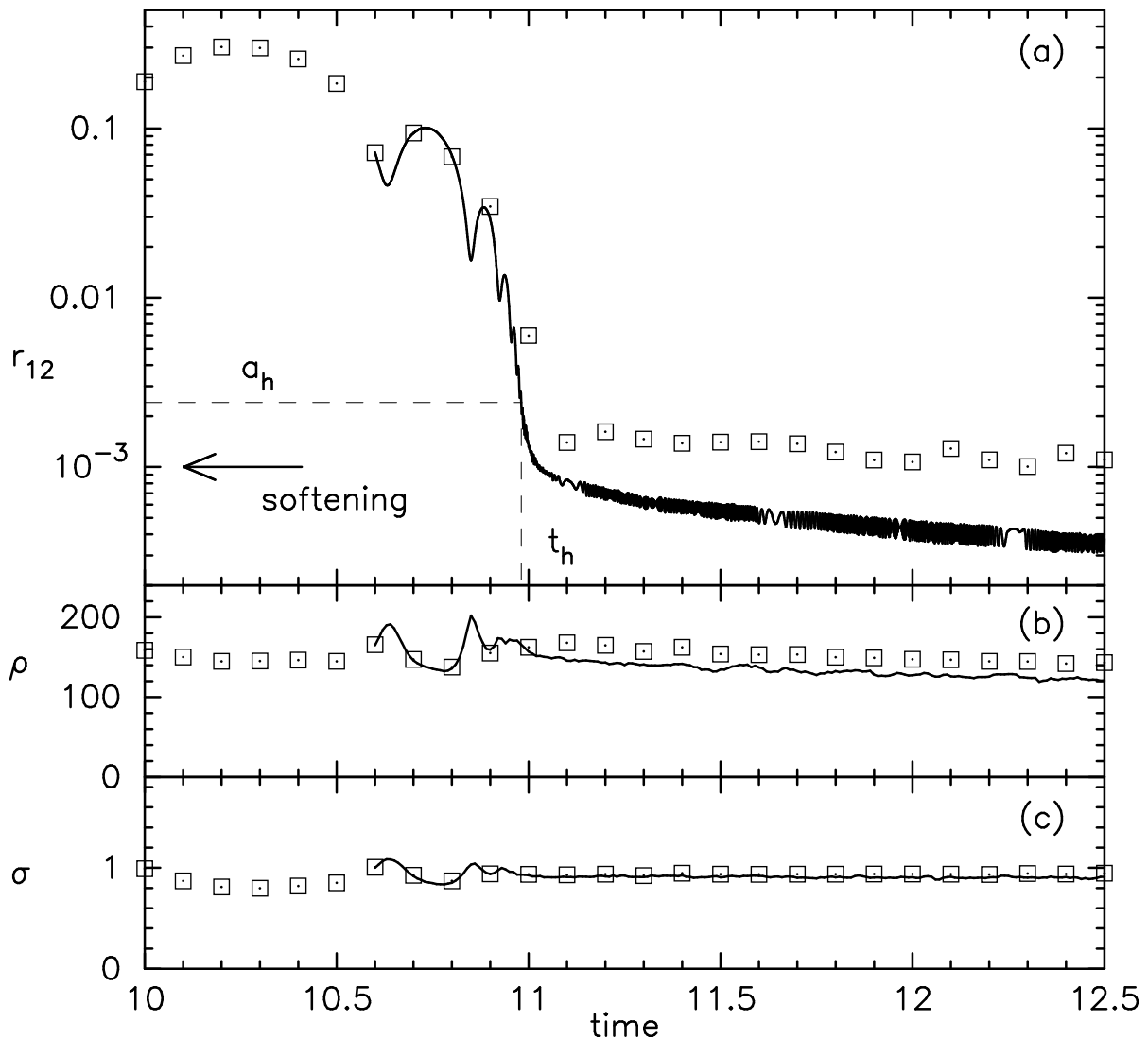}
\epsscale{1.0}
\caption{(a) Evolution of the separation between the BHs.
Time is measured from the start of the simulation in units such that
the crossing time in a single galaxy is $\sim 2.2$.
The BHs form a hard binary at $t\equiv t_h\approx 11.0$.
Squares are from the $2.62\times 10^5$-particle integration with 
the tree code \GDT; 
the binary separation saturates at roughly the softening length
$h$, marked by the arrow, in this simulation. 
Solid line is from the \NB integration {\sf A2} with $M_\bullet/m_*=655$.
\NB is able to follow the decay of the binary to arbitrarily small scales.
(b) Stellar density as a function of time of 
stars separated by distance $r\leq 0.04$ from either of the BHs.
(c) Stellar velocity dispersion within 
$0.01\leq r\leq 0.04$ around each black hole.\label{fig_gdtnb6}}
\end{figure*}

In order to verify the accuracy of the tree code integrations,
we continued the \GDT run for several dynamical times after $t_0$ and
compared the results with the \NB integrations.
Coincidence was found to persist until the separation $r_{12}$ between 
the BHs was $\sim h$, at which point the BHs start to ``see'' 
each other's spurious finite extent in the softened \GDT integrations
(Figure \ref{fig_gdtnb6}).
From that point on, the binary separation saturates with \GDT but 
continues to decrease with \NB.
We conclude that \GDT faithfully reproduces the dynamics of 
merging stellar cusps and BH binaries in regimes where the binary is soft.  

\NB would seem to contain all of the machinery necessary for efficiently 
and accurately handling star-star, BH-BH and BH-star interactions.  
In fact, some of the greatest strengths of \NB in the context of globular 
cluster simulation were found to be weaknesses when the code is applied 
to systems with massive BHs.   Chain regularization becomes impractical when 
some of the bodies are much more massive than others.  
In a model galaxy with $N=10^6$ stars and a dense $\rho\sim r^{-2}$ stellar 
cusp harboring a $M_\bullet/m_*=10^3$ central BH, 
there are $\sim10^3$ bodies whose orbits are largely determined by the 
BH's potential.  
Although these stars satisfy the requirements for KS regularization 
(tiny mutual separations and short time steps), 
only the forces between the BH and the stars are significant; 
star-star encounters are energetically unimportant.  
However \NB is incapable of making this distinction.  
It will either try to generate a KS chain containing $10^3$ bodies, 
which would defeat the purpose of regularization altogether and is 
beyond the present design; or, if one ``turns off'' the chain regularization, 
\NB will resort to a pairwise KS regularization whenever two stars come close, 
even if the motion of both stars is mostly in response to the force from a BH. 
In order to make the integrations go efficiently,
it was sometimes necessary to remove ``by hand'' a few particles
that were in tight orbits around one of the BHs.

Our simulations uniquely incorporate three features. 
1. The galactic merger leading to the formation of the binary BH
is carried out from a state in which both galaxies are reasonably
isolated.
2. Both galaxies initially contain faithful realizations of steep 
stellar density cusps.
3. The BHs are present in the cusps at the outset in a kinematically
consistent manner.  
Earlier studies have incorporated some of these features but never all
of them.
\citet{mae96} reported mergers of King models with 
$N=16,384$ particles and central BHs with $\mh\geq 1/64$.
They conducted repeated mergers by recycling the merger products into 
initial conditions for successive mergers, after 
replacing the pair of BHs by a center-of-mass particle and reducing 
the number of particles by half.
Makino \& Ebisuzaki's choice of King models as initial conditions made
their galaxies poor representations of real stellar spheroids which
always contain power-law density cusps;
nor could they test the hypothesis that weak cusps are generated 
by the interaction of BHs and surrounding stars.
\citet{mak97} studied the evolution of a BH binary produced by a 
similar sequence of mergers, also using King-model initial conditions, 
but with a much larger maximum number of particles, $N=262,144$.  
While the number of particles in Makino's and our simulations is 
effectively the same, judging from the density profiles in Figure 4 of 
his paper, Makino's initial conditions appear to be 
$\sim10^2$ times less dense than ours inside 
of the binary's sphere of influence.
\citet{quh97} studied the evolution of a BH binary 
inside cuspy models with $\rho\sim r^{-1}$ and $\rho\sim r^{-2}$
and a wide range of BH masses 
and particle numbers, $N\leq 2\times10^5$.  
As discussed above,  Quinlan's code was unable to simulate an 
actual merger due to the limitations of its treatment of the
mean field.
All of the detailed results in their paper were derived 
from initial conditions consisting of a {\it single} galaxy into which 
two ``naked'' BHs were dropped from starting points located diametrically 
apart at the half-mass radius.
This configuration is likely to produce substantial evolution of the cusp 
{\it before} the formation of the binary as the infalling BHs heat the stars;
this can in fact be seen in their Figure 1, a plot of Lagrangian
radii over time.
As in the simulations of \citet{mae96} and \citet{mak97},
the initial density of stars around the BH binary in the simulation of
\citet{quh97} was much lower than in ours.
This difference will turn out to be consequential (\S 4).

\citet{bar99}  presented simulations of mergers of identical galaxies
with power-law cusps and no BHs.
He used $N=65,536$ bodies and a fixed time step, leap-frog scheme.
Barnes showed that, outside of the softening length, 
power-law cusps as steep as $\gamma=2$ are preserved by the merger.
We find an analogous result (\S\ref{sec_merge}).

\begin{figure*}[t]
\epsscale{1.6}
\plotone{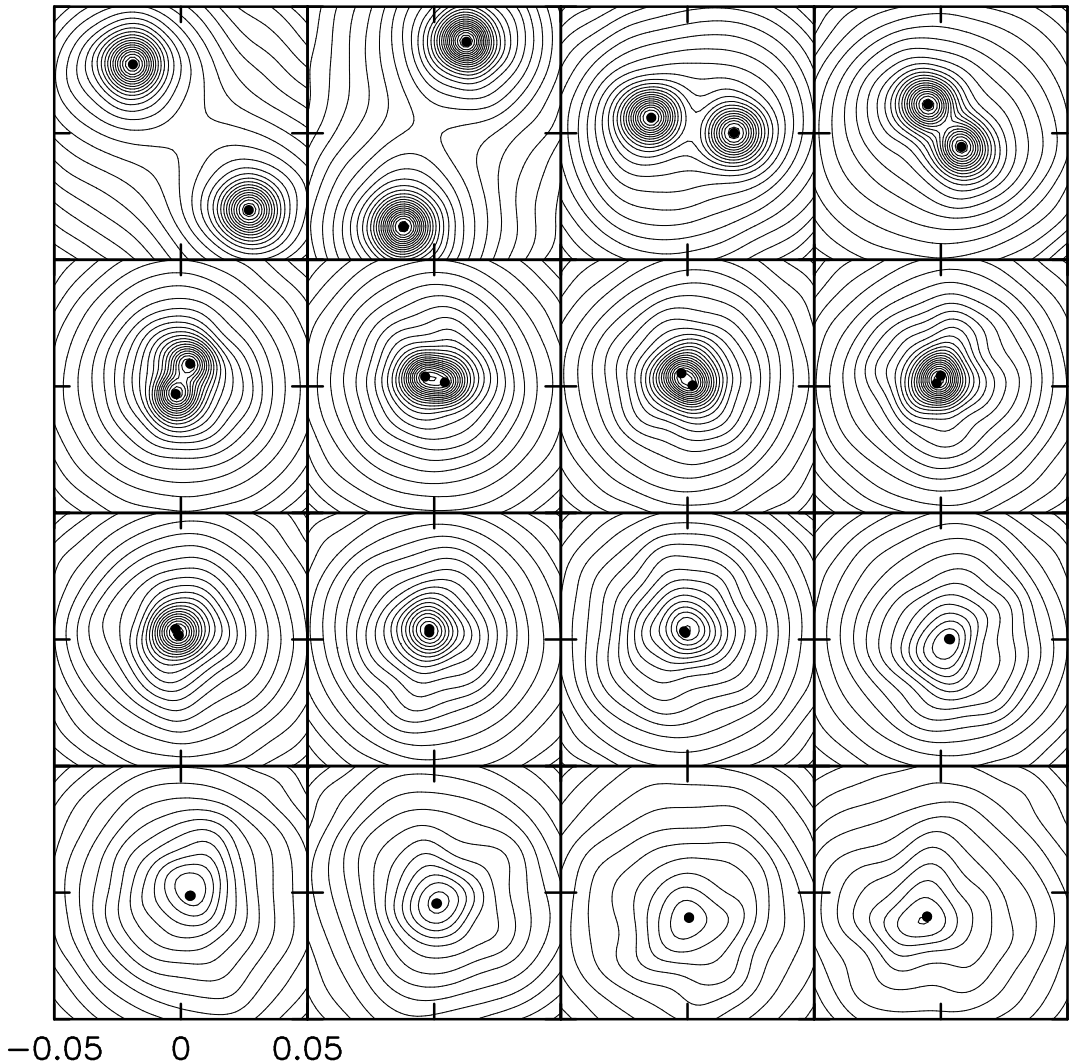}
\epsscale{1.0}
\caption{
Projected density contours for the run {\sf A2} (\NB) with 
$M_\bullet/m_*=655$.
The orbital motion of the BHs (positions indicated by filled circles)
is clockwise in the plane of the figure.
First row: $t= 10.67, 10.8, 10.98, 10.91$.
The two cusps spiral-in under the influence of dynamical friction.
Second row: $t= 10.94, 10.95, 10.96, 10.97$.
The two cusps merge into one.
The final density profile is similar to that of the initial stellar systems.
Third row: $t= 10.98, 11.0, 11.1, 11.6$.
The density of the newly-formed cusp drops rapidly as the BH binary
transfers energy to the stars.
Fourth row: $t= 12.6, 13.6, 16.6, 18.6$.
Density continues to drop as the BH binary ejects stars.
\label{fig_cont}}
\end{figure*}

Our simulations, being purely dynamical, preclude any non-dynamical
processes such as gas-driven dissipation that might act to accelerate
the binary's decay.
Decay might also be enhanced by dynamical processes that
we do not include, e.g. the passage of a star cluster,
gas cloud or third supermassive BH through the nucleus.
How different is the signature on the stellar distribution
of a binary that coalesces immediately after its formation?
We address this question by shadowing each simulation introduced
above with another where the BH binary is replaced by a single BH 
of mass $2\mh$ at time $t=t_0+0.4=t_h$.
This yielded a second set of initial conditions labelled
as ``{\sf Bn}'' in Table \ref{tab_runs}.

\section{Cusp Coalescence}
\label{sec_merge}

We divide the evolution into two regimes, 
before and after the formation of a hard BH binary,
and discuss the first regime in this section.
The two regimes correspond approximately, but not exactly,
to the intervals before and after the start of the 
\NB integrations; as discussed above, these integrations
were begun when the BHs were still a few softening lengths apart
and had not yet formed a tightly bound pair.

As \citet{qui96} notes, there are many ways to define a ``hard'' binary.
The standard definition is energy based:
a binary is hard if its binding energy exceeds the typical particle kinetic
energy, $|E_b| \gg 3m_*\sigma^2/2$.
This definition is inapplicable to the case of massive BH binaries 
since the binary-to-stellar mass ratio scales with $N$ while 
$\sigma$ is independent of $N$; according to this definition, 
a very massive BH binary would always be hard if bound and soft otherwise.  
The famous law of \citet{heg75}, asserting that hard binaries evolve toward 
even harder states, suggests a second definition of hardness.
While a viable distinguishing criterion for hard binaries in star clusters, 
Heggie's law fails to capture the transition between two different processes
---dynamical friction and mass ejection---that {\it both} tend to drive a 
massive binary to an ever-harder state in our simulations.

We therefore followed the suggestion of \citet{hil83} and \citet{qui96}
and defined a ``hard'' binary in terms of its orbital velocity.
The orbital velocity of each BH in a circular-orbit binary is 
$v_c^2=G\mbin/4a$ with 
$\mbin$ the combined mass of the two BHs and $a$ their separation.
We defined the critical separation at which a binary becomes ``hard'' as
\beq
a_h={G\mbin\over 8\sigma_*^2}
\label{def_hard}
\eeq
corresponding to $v_c=\sqrt{2}\sigma_*$.
In model units, $a_h\approx 2.5\times 10^{-3}$ and the binary
separation first falls below $a_h$ at $t\equiv t_h\approx 11.0$
(Figure \ref{fig_gdtnb6}).
``Subsonic'' massive binaries, $a\gg a_h$, 
harden by dynamical friction acting on each BH (and its associated
cluster) individually;
a ``supersonic,'' or hard, binary behaves like a structureless point mass 
under the action of dynamical friction but can capture stars and eject 
them at much higher velocity, thereby increasing its hardness.
\citet{qui96} notes that this definition of hardness is roughly equivalent
to the statement that a hard binary hardens at a constant rate.
We found this to be true (\S 4); the definition (\ref{def_hard})
is also a natural one in the sense that the character of the binary's 
evolution, 
and the evolution of its surrounding stellar cluster,
were found to undergo qualitative changes when $a$ dropped below $\sim a_h$,
corresponding to the onset of mass ejection by the binary.

The merging of the stellar cusps for run {\sf A2}
is illustrated in Figure \ref{fig_cont}.
This figure makes manifest that the BHs remain closely associated with
their initial stellar cusps during every stage of the merger, 
up to and including the point when the two cusps merge into one 
at $t\approx t_h$.
A consequence is that {\it dynamical friction brings the BHs together
much more rapidly than if they were ``naked,'' since their effective
mass is the mass of the cluster of stars bound to them}.

We can check this assertion by comparing the orbital decay rates for an 
isolated BH and that embedded in a stellar cusp.  
The decay rate for an isolated BH on an approximately circular orbit is given by 
\citep{bit87}
\begin{eqnarray}
\frac{da}{dt}&=&-\frac{\erf(1)-\erfp(1)}{\sqrt{2}} \frac{GM_\bullet}{\sigma_* a} \ln\Lambda\nonumber\\
&\approx& -0.302 \frac{GM_\bullet}{\sigma_* a} \ln\Lambda
\end{eqnarray}
where $a$ denotes separation between two BHs.  
With $M_\bullet=0.01$, $a=0.1$ and $\sigma_*\approx 2^{-1/2}$, the decay rate 
is estimated at $da/dt\approx -0.043\ln\Lambda$.  
Note that for $a>0.1$ the formula yields
even lower rate of decay.  As for the Coulomb logarithm, it can be written
in terms of the ratio of the maximum and the minimum impact parameter 
$\ln\Lambda=\ln(p_{max}/p_{min})$ where it is a standard choice to 
select the gravitational radius of the isolated ``test particle'' 
for the latter, $p_{min}=GM_\bullet/\sigma_*^2$.
In lack of a canonical choice for $p_{max}$, we equate it to 
the orbital radius, which implies 
$\ln\Lambda=\ln5.0\approx 1.6$ and thus $da/dt\approx-0.07$ (but see also Appe/ndix A).  
The predicted decay rate is a factor of $\sim6$
smaller than the rate of in-spiral $da/dt\approx -0.43$ 
we measured in the \GDT run in the interval $9.0\leq t\leq 11.0$ preceding 
the formation of hard binary, in this interval
$a(t)=0.78-0.43\times(t-9.0)$ is a good fit.

We compare the isolated particle estimate with an estimate of how rapidly 
dynamical friction would act to bring together two overlapping spheres 
with $\rho\sim r^{-2}$ density profiles. 
Let the spheres each have mass ${\cal M}$ and density 
$\rho=\sigma_*^2/2\pi G r^2$ inside a radius $a$. 
When the separation $a$ between their centers is much larger than the BH radius of influence, or equivalently ${\cal M}(a)\gg M_\bullet$, we can ignore the BHs. 
Then ${\cal M}(a) = \sigma_*^2a/G$ and the circular velocity is
$v_c^2(a)=\sigma_*^2/2$.
The dynamical friction force acting on one of the spheres is given 
by
\begin{eqnarray}
\langle\Delta v_{\parallel}\rangle &=& -{4\pi G^2{\cal M}\rho\ln\Lambda F(v)\over v^2} , \nonumber\\
F(v) &=& \erf(x) - x\erfp(x)
\label{eq_chandra}
\end{eqnarray}
\citep{cha43}, where $\ln\Lambda$ is the Coulomb logarithm and
$x=v/\sqrt{2}\sigma_*$.
Setting $v=v_c=\sigma_*/\sqrt{2}$ gives $F=0.0811$.
Taking for $\rho$ the density of either sphere at a distance
$a$ from its center, we find
\beq
\langle\Delta v_{\parallel}\rangle \approx -{0.324\sigma_*^2\ln\Lambda\over a}.
\eeq
Equating the torque produced by this acceleration with the change
of orbital angular momentum $J$, and writing
\beq
{dJ\over dt} = {dJ\over da}{da\over dt} = {\sigma_*^3 a\over \sqrt{2}G}{da\over dt},
\eeq
gives
\beq
{da\over dt} \approx -0.23\sigma_*\ln\Lambda.
\eeq
Just outside of the BHs' sphere of influence, $\sigma_*\approx 1$
(Figure \ref{fig_gdtnb6}b).
For $\ln\Lambda$ we again take $\sim 1.0$,
giving $da/dt\approx 0.24$.  
This result still falls short of the observed value 
($da/dt\approx-0.43$) by a factor of $\sim 2$.
  
A detailed integration (Appendix A) taking into account the shape and 
the finite extent of both spheres yields the rate
$\langle\Delta v_{\parallel}\rangle\approx -1.50\sigma_*^2/a$ (note the 
absence of $\ln\Lambda$), therefore $da/dt\approx -1.06$.  This result, however,
is sensitive to the choice of tidal radius outside of which the spheres 
are indistinguishable; estimation of this radius is difficult in part 
due the large orbital eccentricity of the galaxies in the simulation.

\begin{figure}
\plotone{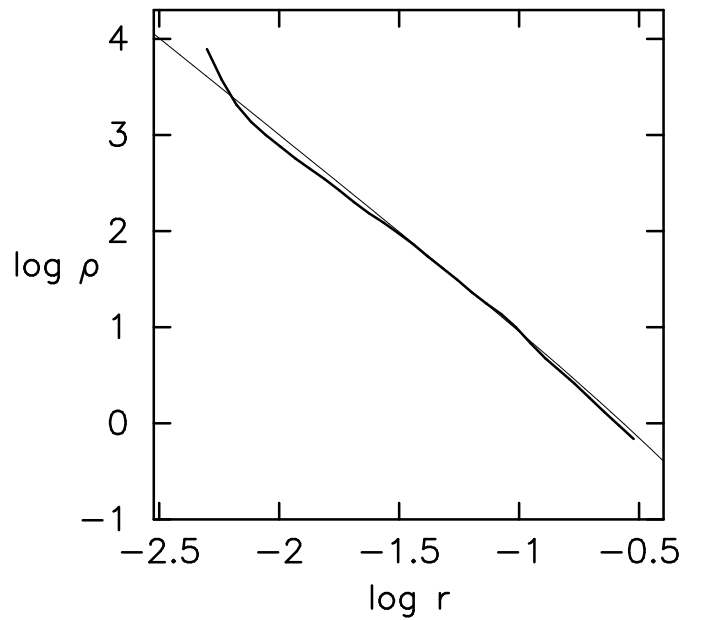}
\caption{
Radial density profiles of the pre-merger galaxies (thin curve) 
and at time $t\approx t_h=10.96$ (thick curve) when the binary 
separation equals $\rbin=6.5\times10^{-3}$.  
The pre-merger density was multiplied by the factor 
$(Mr_0^{-3})_{new}/(Mr_0^{-3})_{old}\approx 0.53$  
to bring it to the scale of the post-merger galaxy.
The two galaxies have merged into a single galaxy that is nearly
homologous with the initial galaxies on scales $r\gtrsim \rbin/2$.
Shortly after this time, the central density drops as the binary 
heats the core.\label{fig_homol}}
\end{figure}

\begin{figure}[t]
\plotone{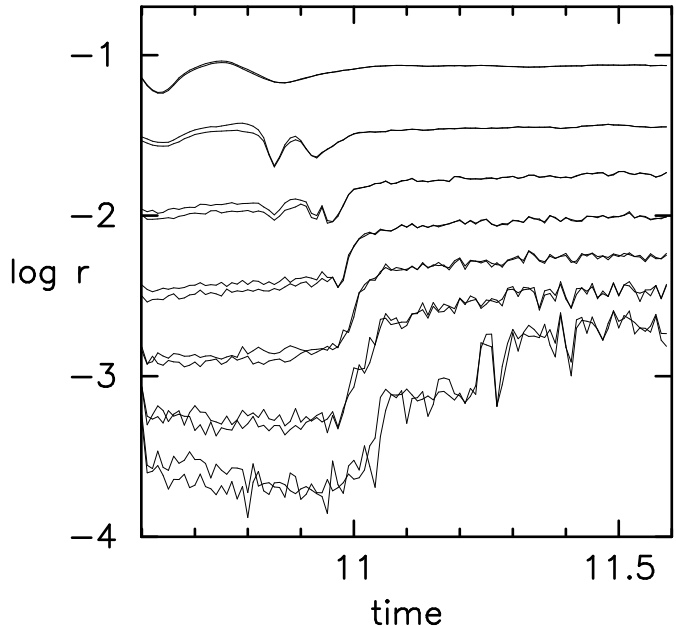}
\caption{
Lagrangian radii around each BH in the first time unit of the
\NB run {\sf A2}.
From bottom to top, the radii enclose $10^{-4}$, $10^{-3.5}$, $10^{-3}$, $10^{-2.5}$, 
$10^{-2}$, $10^{-1.5}$ and $10^{-1}$ in units of the mass of one galaxy before the merger.\label{fig_lagrange}}
\end{figure}

Remarkably, the density structure of the merged galaxy {\it just
after} formation of a hard binary is essentially identical to 
that of the initial stellar systems at radii $r\gtrsim a$.
This is illustrated in Figure \ref{fig_homol}.
Homology following a merger was found also by \citet{bar99}
in spherical galaxies without BHs.
In our simulations, however, the homology is short-lived.
The formation of a hard BH binary at $t\approx t_h$ is followed 
by a sudden drop in the stellar density within the binary's 
gravitational sphere of influence, $r\lesssim 0.01$.
This is clearly seen in Figure \ref{fig_lagrange}, a plot
of Lagrangian radii, and also in the density contour plot of
Figure \ref{fig_cont}.
(The drop in density is not so apparent in Figure \ref{fig_gdtnb6}b 
because the density plotted there is an average over a radius of
$0.04$, and little net change in density occurs within this radius 
-- see Figure \ref{fig_lagrange}.)
In effect, the steep cusp that was present immediately after formation
of the BH binary is destroyed in little more than the local crossing time.

What is responsible for the rapid destruction of the cusp?
Two possible, and closely related, mechanisms are deposition of 
energy into the stars by dynamical friction acting on the BHs
individually; 
and ejection of stars that exchange energy with the BH binary.
Neither process is well defined in this regime where the
BH binary is neither very hard nor very soft.
Nevertheless we can write approximate expressions for the rate
at which energy is transferred to the stars by the two mechanisms,
by assuming either 
that the BHs are moving independently of each other, 
or as members of a tight binary;
of course neither assumption is strictly satisfied.

In the first case, dynamical friction would extract energy from the 
two BHs at a rate
\beq
\langle\Delta E\rangle = 2Mv\langle\Delta v_{\parallel}\rangle = 
-{8\pi G^2M^2\rho\ln\Lambda F(v)\over v}
\eeq
(cf. equation \ref{eq_chandra}).
Setting $v=\sqrt{2}\sigma_*$, our definition for the onset of a
hard binary, gives $F=0.43\approx 1/2$ and
\beq
\langle\Delta E\rangle \approx -2\sqrt{2}\pi G^2M^2\rho\sigma_*^{-1}\ln\Lambda .
\eeq
The alternate mechanism, 
hardening of the binary by mass ejection, produces energy at the rate
\beq
{dE\over dt} = {G^2M^2\rho H\over 2\sigma_*}
\eeq
\citep{hil83,miv92,qui96},
where $H$ is the dimensionless hardening rate;
$H\approx 15$ in the limit of a very hard, equal-mass binary and drops to 
$\sim 10$ for a binary with $a=a_h$.
Thus, {\it both} mechanisms predict an energy deposition rate that can be written as
\beq
\left|{dE\over dt}\right| = C G^2M^2\rho\sigma_*^{-1}
\eeq
with $C\approx 5$; we have taken $\ln\Lambda \approx 0.5$
(Appendix A).

The energy of the binary when $a=a_h$ is 
$E=-2M\sigma_*^2$, so the characteristic time over which either process
extracts energy is $\sim 2C^{-1}\sigma_*^3/G^2M\rho$.
Computing $\rho$ using the mass within $r=0.01$ at $t=t_h$ gives
an energy extraction time scale of $\sim 0.2$ in model units.
This is quite comparable to the time associated with the jump in
Lagrangian radii of Figure \ref{fig_lagrange}.
We note also that the energy extracted from the binary in this time, 
$E\approx 2M\sigma_*^2\approx 0.02$, 
is comparable to the energy in stars within a radius of $\sim 0.01$.
This too is consistent with the changes in Lagrangian radii shown in
Figure \ref{fig_lagrange}.

\begin{figure}
\plotone{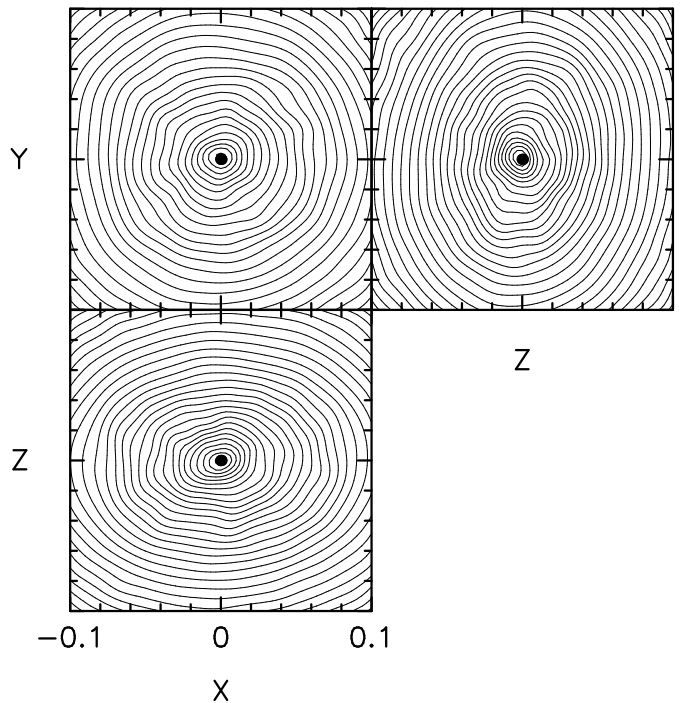}
\caption{
Isophotes in the run {\sf A2} in three projections.  
Black dots show the location of the BH binary; separation between the BHs at 
this time is $a\sim 1.5\times 10^{-4}$. 
100 snapshots of the nucleus were superposed in the interval 
$18.1\leq t<19.1$.  
The merger remnant is approximately axially symmetric with an 
edge-on ellipticity of $\epsilon\approx0.25$.
\label{fig_isophotes}}
\end{figure}

\begin{figure*}
\epsscale{1.6}
\plotone{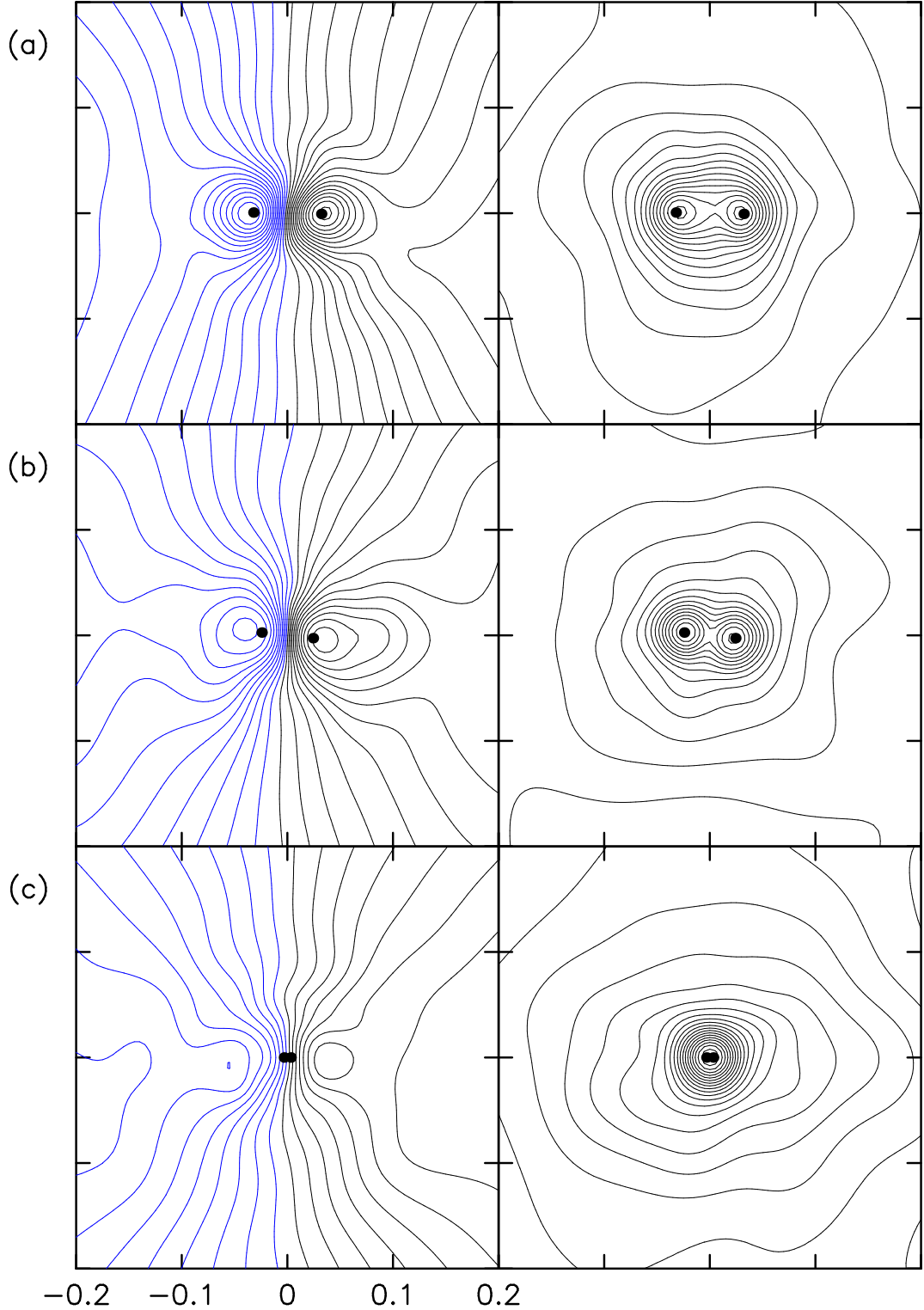}
\epsscale{1.0}
\caption{
2D kinematics of the merging cusps for the run {\sf A2}. 
View is in the plane of the merger from a direction perpendicular to 
the line connecting the two BHs.
Left panels show the mean line-of-sight velocity;
blue contours indicate approaching stars.
Right panels are the line-of-sight velocity dispersion.  
In all panels the contours are separated by $0.038$.
(a) $t=10.66$; (b) $t=10.82$; (c) $t=10.96$.
The BHs remain centered on the velocity dispersion peaks but
move inward with respect to the peak of the rotational velocity.
\label{fig_2dkin}}
\end{figure*}

\begin{figure*}
\epsscale{1.6}
\plotone{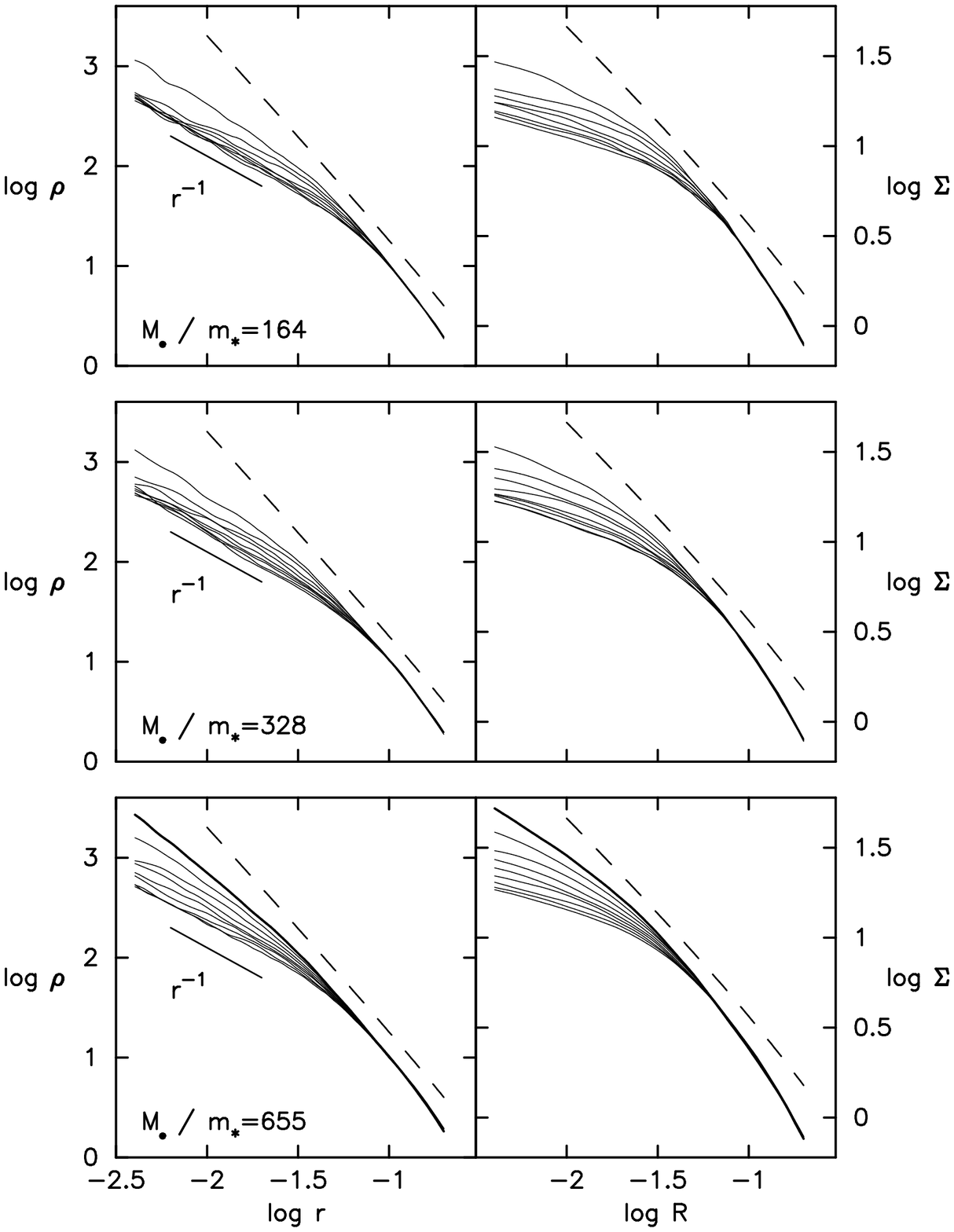}
\epsscale{1.0}
\caption{
Spatial density profiles (left column) and projected density 
profiles (right column). 
In each panel, starting at time $t=11.0$, the profiles are
recorded from top to bottom at intervals $\Delta t=1.0$,
 each of them an average obtained by superposing 100 snapshots.  
Top row, run {\sf A8};
middle row, run {\sf A4}; 
bottom row, run {\sf A2}.
Thick lines represent the run {\sf B2} with one BH; the merger remnant would 
have this profile if the BHs coalesced at $t=t_h$.  
Dashed lines are profiles of the original galaxies 
multiplied by an arbitrary factor.
The merger remnant has a $\rho\sim r^{-1}$ density cusp which
projects to a core profile with continuous curvature.
\label{fig_profs}}
\end{figure*}

We conclude that the sudden disruption of the steep cusp is
attributable to transfer of energy from the BHs into the
surrounding stars as the BHs form
a hard binary.
We emphasize again the {\it rapidity} of this process, which earlier
analyses have overlooked.
If a galaxy's cusp is to avoid this fate, 
some mechanism must extract energy from the binary BH on a time
comparable to the local dynamical time, before it is able to exchange
energy with the stars.

The density near the BHs continues to drop at later times
although more slowly, as the BH binary gradually decays.
We discuss this process in more detail below.

The large-scale kinematical evolution of the merger is illustrated in Figure
\ref{fig_2dkin}.
The general character of these plots is similar to what is seen
in simulated mergers of equal-mass galaxies without BHs (e.g. \citet{beb00}), 
with the highest mean rotational velocities in the plane of the
merger and a roughly cylindrical rotation pattern elsewhere.
The peaks in the rotational velocity initially correspond with the
locations of the BHs, but dynamical friction causes the BHs to move
inward from these peaks at a time $t\approx 10.75$.
Velocity dispersions, on the other hand, remain peaked on the BHs at 
all times consistent with the fact that the BHs remain centered on
their cusps (Figure \ref{fig_cont}).
At the time of formation of the hard BH binary, $t=t_h\approx 11.0$,
the merger remnant is mildly rotating with a peak line-of-sight
rotational velocity of $v/\sigma_*\approx 0.58$ at a distance 
$\sim 0.04$ from the center (using $\sigma_*\approx 0.6$ at distance $0.2$), although bulk fluctuations at $20\%$ level
in the rotation field persist on scales $r\lesssim 0.1$.
The ellipticity of isophotes shown in Figure \ref{fig_isophotes} is $\epsilon\approx 0.25$ which falls on the isotropic oblate rotator relation $v/\sigma_*=\sqrt{\epsilon/(1-\epsilon)}$.

Evolution of the stellar density profiles is shown in Figure \ref{fig_profs}.
Profiles were computed from the $N$-body positions using a 
nonparametric kernel routine based on the algorithms in \citet{met94};
details are given in Appendix B.
Each profile is an average over several snapshots;
they are separated by $\Delta t = 1.0$ starting at $t=t_h=11.0$.
The evolution of $\rho(r)$ becomes more regular as $N$ is increased,
due probably to the smaller random motion of the binary for larger $N$.
Considerable evolution occurs in the interval $t_h-1\lesssim t\lesssim t_h+1$
when the BHs form a hard binary, as discussed above.
A break appears in the profiles at $t\approx t_h$ where the outer,
$\rho\sim r^{-2}$ profile turns over to a shallower inner dependence;
the inner profile is well approximated as a power-law as well, with slope
$d\log\rho/d\log r\approx -1$ that gradually decreases with time.
In projection, this weak power-law cusp produces a core-like profile
with continuously varying slope.
Hence this galaxy would be classified as a ``core galaxy'' for $t\gtrsim t_h$
\citep{lau95}; we note that core galaxies also show weak power-law cusps
on deprojection \citep{mef96}.
We defined the ``break radius'' $R_b$ as the radius where the second
derivative of $\Sigma(R)$ on a log-log plot reaches a minimum;
this definition is consistent with the more common one based on
fitting of a parametric form to the surface brightness profile 
(e.g. \citet{lau95}).
We defined $r_b$ in the same way, as the break radius corresponding to
the space density profile $\rho(r)$.
Values of $R_b$ and $r_b$ at several different times are given in Table 
\ref{tab_summary}.

Our simulations are the first to demonstrate that weak, power-law cusps --
corresponding to what are commonly called ``core'' or ``cuspy-core'' 
galaxies -- can be generated by the merger of galaxies with steep cusps,
or ``power-law'' galaxies.
Since core galaxies are systematically brighter than power-law galaxies, 
it is natural to suppose that weak cusps have
their origin in mergers.
We will explore this hypothesis in more detail below, 
after discussing the further evolution of the density profiles
that takes place as the BH binary slowly decays.
Here we discuss one problem with the hypothesis, and a possible resolution.
Even some moderately bright elliptical galaxies ($M_V\approx -22$) exhibit
steep cusps, even though these galaxies have certainly experienced mergers 
in the past.
How did these galaxies avoid the rapid cusp destruction that takes
place in our simulations?

A possible answer is suggested by Figures \ref{fig_homol} and 
\ref{fig_profs}.
Immediately after the merger, at $t\approx t_h$, the density profile
is briefly almost homologous with the initial profile,
with a steep, $\rho\sim r^{-2}$ cusp.
If some mechanism could induce a rapid coalescence of the BHs at this time,
before the BH binary was able to exchange energy with the stars,
the steep cusp might avoid disruption.
We tested this idea using our runs in which the two BHs were artificially
combined into one at $t=t_0+0.4=11.0$.
The test was successful; the density profile after coalescence of the binary
(shown as the heavy line in the bottom panels of Figure \ref{fig_profs})
is indeed very close to the initial profile and remains so indefinitely.
We discuss below (\S\ref{sec_discuss}) whether any mechanism might exist
for inducing such a rapid coalescence, and why it should preferentially
be active in low-luminosity galaxies.

\section{Evolution of the Black-Hole Binary and its Effect on the Structure of the Nucleus}
\label{sec_hard}

\subsection{Physical Processes}

The evolution of a massive BH binary in a galactic nucleus has
been discussed by a number of authors.
We begin by summarizing that work here and listing the
physical processes that govern the evolution of the binary
and its effect on the surrounding stars.

1. {\it Hardening of the binary.}
Stars that pass within a distance $\sim a$ of the binary,
with $a$ the binary's semimajor axis, experience a gravitational slingshot
and are ejected with velocities 
$v_{ej}\approx V_{bin}\equiv\sqrt{G\mbin/a}$
(e.g. \citet{hif80}); $V_{bin}$ is the relative velocity
of the two BHs if their orbit is circular and $\mbin=M_1+M_2$ is the
total mass of the binary.
In a fixed stellar background, this leads to hardening at a rate
\begin{equation}
{d\over dt}\left({1\over a}\right) = {G\rho\over\sigma_*}H
\label{defhard}
\end{equation}
and the rate of energy extraction from the binary is
\beq
{dE_b\over dt} = {G^2\mbin^2\rho\over 8\sigma_*}H
\label{defextract}
\eeq
with $H$ a dimensionless hardening rate.
Here $\rho$ and $\sigma_*$ are the mass density and 1D velocity dispersion
of the stars.
Equations (\ref{defhard}) and (\ref{defextract}) 
are derived from a model in which the stellar density is
assumed uniform and the gravitational field from the stars is
ignored; gravitational focusing by the BH binary is incorporated
but not the influence of the stellar potential on stellar orbits.
The dimensionless hardening rate $H$ is a function of the hardness
of the binary, measured for instance by $V_{bin}/\sigma_*$,
as well as the binary's mass ratio $M_1/M_2$ and eccentricity.
For an equal-mass, circular-orbit binary, $H$ varies from 
$\sim 15$ for an infinitely hard binary to 
$\sim 2.0$ for $V_{bin}/\sigma_*=1$ \citep{qui96}.

Stellar encounters also modify the binary's  orbital eccentricity $e$.
The eccentricity growth rate,
\beq
\label{eq_eccgrow}
K={de\over d\ln(1/a)},
\eeq
is negligible for $V_{bin}/\sigma\approx1$ and increases to a maximum of $\sim 0.2$ for an 
equal-mass binary with $e\approx 0.7$ and $V_{bin}/\sigma\gtrsim 20$
\citep{miv92,qui96}.
Changes in eccentricity are potentially important because the 
rate of orbital energy loss due to gravitational radiation grows steeply 
for $e\rightarrow 1$ (cf. equation \ref{eq_tgrav}), 
hence an eccentric binary will coalesce sooner than a circular one
with the same $a$.

2. {\it Mass ejection.}
The binary ejects mass at a rate
\begin{equation}
J = {1\over\mbin}{dM_{ej}\over d\ln(1/a)} 
\label{defeject}
\end{equation}
where $J\approx 1$ is nearly independent of ($M_1/M_2$, $a$) 
for $a\ll a_h$ and drops with decreasing hardness of the binary
\citep{miv92,qui96}.
Ignoring the variation of $J$ with $V_{bin}$, 
one can integrate equation (\ref{defeject}) to obtain
\begin{equation}
M_{ej} \approx J M_{12} \ln(a_{ej}/a)
\label{eq_mejlaw}
\end{equation}
where it has been assumed that mass ejection begins when 
$a=a_{ej}$; we expect $a_{ej}\approx a_h$.
Thus the binary ejects of order its own mass in shrinking from
$a= a_{ej}$ to $a=a_{ej}/2$.
If the binary's mass is not negligible compared with the 
mass of the pre-existing nucleus, the stellar density near the binary
will drop as the decay proceeds, 
causing the hardening rate to also drop (equation \ref{defhard}).

3. {\it Brownian motion.}
The binary exhibits Brownian motion due to momentum imparted by
encounters with stars.
A single particle of mass $M$ in statistical equilibrium with a 
Maxwellian field of light scatterers with masses $m_*\ll M$
will exhibit an average speed
determined by equipartition of energy,
\beq
\mbin \langle v^2\rangle = 3m_*\sigma_*^2,
\label{equipart}
\eeq
and its radius of wandering 
$r_w$ will be given by
\beq
\langle r_w^2\rangle \approx {\langle v^2\rangle\over G\rho}
\label{defwander}
\eeq
where $\rho$ is a mean density averaged over the wandering region.
These relations ignore any reaction of the background particles
to the motion of the massive object.
Corrections also apply if the massive object is a binary,
which receives larger kicks than a point mass from ejected stars.
The speedup is at most a factor of $\sim 2$ for a very hard binary in 
a nucleus with a steep density profile \citep{mer01}.
Brownian motion is potentially important
because it allows the binary to interact with
a larger pool of stars than if it were fixed at the center of
the potential, thus prolonging its decay.
However the amplitude of the wandering in an $N$-body simulation
is likely to be much larger than in a real galaxy due to the unphysically
small value of $\mbin/m_*$ in the simulations.
Brownian motion may also help to scatter stars into the binary's
sphere of influence by introducing a complex time dependence into 
the gravitational potential felt by the stars.

4. {\it Loss-cone refilling.}
Eventually the binary will eject most or all of the stars
which can come within a distance $\sim a$ of it.
If the binary wanders over a distance $r_w>a$,
this will happen when it has ejected all stars whose pericenters lie 
within a distance $\sim r_w$ from the galaxy center.
Once the density of stars in the vicinity of the binary
drops to zero, the binary's decay will stall, unless 
some process can refill the ``loss cone.''
One possibility is infall of a third BH,
gas cloud, dwarf galaxy or other massive object that can perturb
the stellar orbits.
In the absence of such dramatic events, 
ordinary two-body relaxation will scatter stars into the binary's
sphere of influence.
The associated feeding rate is
\beq
{dM_{scat}\over dt}\approx {M(r_{max})\over t_r(r_{max})\ln(r_{max}/r_w)}
\label{losscone}
\eeq
where $M(r)$ is the stellar mass within $r$ and $t_r$ is the
star-star relaxation time; $r_{max}$ is the radius at
which the rate of scattering into the loss cone peaks,
typically of order $r_{gr}$ \citep{sha85}.

5. {\it Gravitational radiation.}
If the decay of the binary continues sufficiently far,
emission of gravitational radiation will eventually become
the dominant source of energy loss.
The gravitational radiation time scale $t_{gr}$ is
\beq
t_{gr} = \left|{\dot a\over a}\right|_{gr}^{-1} = 
{5\over 64} {c^5a^4\over G^3M_{12}^3} F(e)
\label{eq_tgrav}
\eeq
where the factor $F(e)$ contains the eccentricity dependence:
\beq
F(e)=\frac{\left(1-e^2\right)^{7/2}}{1+\frac{73}{24}e^2
+\frac{37}{96}e^4}
\label{fofe}
\eeq
\citep{pet64}.
The dependence of $t_{gr}$ on $e$ is weak for small $e$;
$F(0)=1$ and $F(0.5)\approx 0.205$.
The decay rate from gravity wave emission matches that from
stellar ejection when
\beq
a^5=a_{crit}^5 \equiv {64\over 5FH}{G^2\mbin^3\sigma\over c^5\rho}.
\label{acrit}
\eeq
The right hand side of this expression is difficult to evaluate 
{\it ab initio} since
$\rho$ will be strongly affected by stellar ejection
during the binary's decay.
However semi-analytic models for the combined evolution of the binary and
the nucleus \citep{mer00} suggest that
$a_{crit}\approx 10^{-2}a_h$.
Equation (\ref{eq_mejlaw}) then implies that the binary must eject
roughly four times its mass in stars in order to achieve
gravitational radiation coalescence.
We do not include an energy sink term corresponding to gravitational
radiation in our simulations but use equation (\ref{acrit}) to estimate
when coalescence would occur.

A crucial question when interpreting $N$-body simulations
is the dependence of the results on $N$.
Real galaxies have nuclei with $N\approx 10^7$, much greater
than the particle numbers amenable to computer simulation.
Fortunately, the two processes that most directly affect the
evolution of a BH binary in a galactic nucleus -- hardening 
and mass ejection -- have rates that depend only on the local 
{\it density} of stars, not on their masses.
The rates at which these two processes occur in our simulations 
should therefore reflect their rates in galaxies
whose mass distributions are similar to those in our models.

\begin{figure*}[t]
\epsscale{1.8}
\plotone{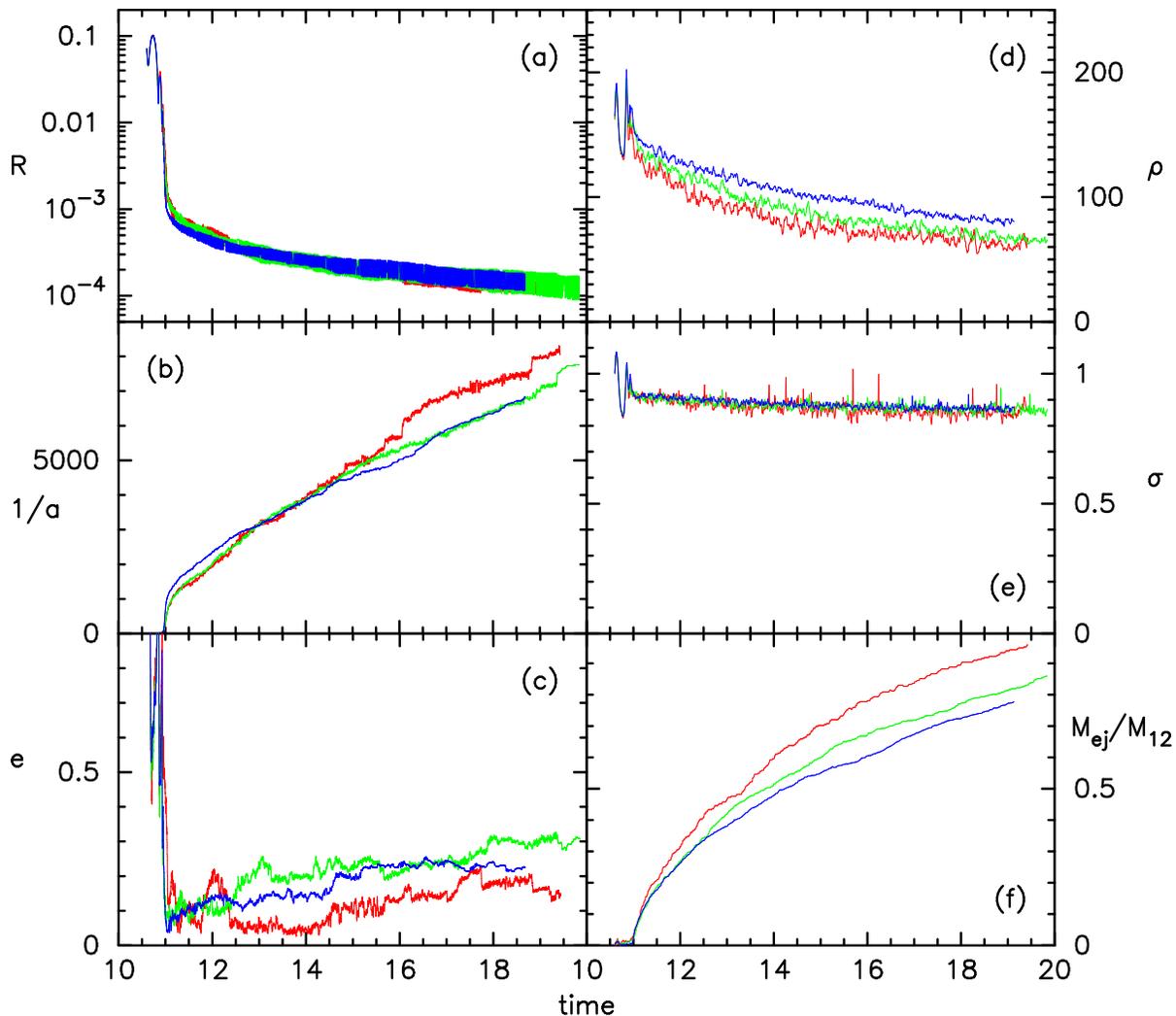}
\epsscale{1.0}
\caption{
Evolution of the BH binary and response of the stellar environment.
Red curves: run {\sf A8} ($M_\bullet/m_*=164$);
green curves: run {\sf A4} ($M_\bullet/m_*=328$); 
blue curves: run {\sf A2} ($M_\bullet/m_*=655$).  
(a) Separation between the BHs; 
(b) inverse semimajor axis $a^{-1}\equiv-2E_b/GM_\bullet^2$;
(c) orbital eccentricity.
When calculating $a$ and $e$ before $t_h$, we corrected the BH masses
by adding the mass in stars bound to each of them.
(d) Average density of stars within $r\leq 0.04$ from either of the BHs; 
(e) velocity dispersion in a region $0.01\leq r\leq 0.04$ 
from either of the BHs;
(f) total mass of stars ejected by the binary, in units of the
binary mass.\label{fig_binary}}
\end{figure*}

However both the Brownian motion of the binary
and the refilling of the binary's loss cone by two-body encounters
are $N$-dependent processes, and their importance in our simulations
is expected to be much greater than in real galaxies.
The binary's wandering radius scales as
\beq
r_w\propto \sqrt{m_*\over\mbin} \propto N^{-1/2}
\eeq 
while the rate of scattering of stars into the loss cone varies as
\beq
{dM_{scat}\over dt}\propto t_r^{-1}\propto m_*\propto N^{-1}
\eeq
where the $N$-dependence of the Coulomb logarithm has been ignored.
These ``second order'' effects do not directly influence the binary's
evolution but they do determine how large a supply of stars
is available to the binary and hence how long its orbit can
continue to decay.
For instance, a wandering binary can interact with a larger pool of 
stars than a binary that is stationary.

In a real galactic nucleus where $N$ is very large, 
Brownian motion and two-body relaxation would be expected to
be almost negligible.
(Exceptions might occur in nuclei where the gravitational
potential is very lumpy, due to giant molecular clouds,
star clusters, additional massive BHs etc.)
An obvious inference, drawn by several authors 
\citep{val96,mer00,zie00,gor00},
is that a massive BH binary should rapidly eject those stars 
whose orbits bring them within its sphere of influence,
after which the binary separation should cease to change.
These arguments are not air-tight however because of the complicated
way in which the various physical processes interact.
For instance, mass ejection lowers the density of stars, causing
the hardening rate to drop (equation \ref{defhard}),
but the reduction in density leads to an increase in the wandering
radius (equation \ref{defwander}) allowing the binary to move out of the
low-density region into a region of higher density where it can continue 
interacting with stars.

\subsection{Hardening rate and mass ejection}

\begin{deluxetable}{ccrrr}
\tablecolumns{5} 
\tablewidth{19pc}
\tablehead{
\colhead{Quantity} &
\colhead{Time} &
\colhead{{\sf A8}} & 
\colhead{{\sf A4}} & 
\colhead{{\sf A2}}}
\tablecaption{Measured $N$-Body Parameters\label{tab_summary}}
\startdata
$r_b$ & $11$ & 0.038 & 0.036 & 0.035 \\
$r_b$ & $15$ & 0.049 & 0.048 & 0.035 \\
$r_b$ & $19$ & 0.087 & 0.086 & 0.056 \\
$R_b$ & $11$ & 0.036 & 0.033 & 0.026 \\
$R_b$ & $15$ & 0.053 & 0.068 & 0.030 \\
$R_b$ & $19$ & 0.080 & 0.072 & 0.063 \\
$da^{-1}/dt$ & $[11,14]$ & 970 & 920 & 750 \\
$da^{-1}/dt$ & $[16,19]$ & 660 & 540 & 690 \\
$H$ & $13$ & 9.3 & 7.9 & 5.8 \\
$H$ & $17$ & 8.1 & 6.3 & 6.8 \\
$K$ & $>11.6$ & 0.70 & 0.13 & 0.13 \\
$a_{ej}$ & $>12$ & 0.00098 & 0.00085 & 0.00074 \\
$a_{ej}$ & $>13$ & 0.00085 & 0.00074 & 0.00070 \\
$a_{ej}$ & $>14$ & 0.00090 & 0.00071 & 0.00070 \\
$J_\infty$ & $>12$ & 0.45 & 0.45 & 0.46 \\
$J_\infty$ & $>13$ & 0.49 & 0.49 & 0.48 \\
$J_\infty$ & $>14$ & 0.48 & 0.51 & 0.48 \\
$\sigma_{12}$ & $13$ & 0.087 & 0.064 & 0.034 \\
$\sigma_{12}$ & $16$ & 0.085 & 0.061 & 0.032 \\
$\sigma_{12}$ & $19$ & 0.081 & 0.058 & 0.030 \\
$r_w$ & $[11,19]$ & 0.028 & 0.011 & 0.0084 
\enddata
\end{deluxetable}

Figure \ref{fig_binary} summarizes the evolution of the BH binary in our
simulations.
At any given time, the binary can be described by its semimajor 
axis $a$, its eccentricity $e$, 
the direction of its orbital angular momentum vector ${\bf \hat n}$, 
and the position and velocity of its center of mass.
The binary's hardening rate $da^{-1}/dt$ (Figure \ref{fig_binary}b)
is nearly constant with time following the ``knee'' at $t\approx t_h$
when the binary first becomes hard.
Minute fluctuations in $1/a$ reflect perturbations in the binary's
binding energy due to stars tightly bound to one of the BHs;  
sudden jumps indicate times when the binary ejects a single star.
At low $N$, the discreteness of individual ejection events induces statistical 
fluctuations in the value of $1/a$ that in our opinion are responsible for 
most of the $\sim 20\%$ differences in $1/a$ between runs with different $N$.  
Table \ref{tab_summary} gives values of $da^{-1}/dt$ obtained by fitting 
straight lines to $1/a$ over intervals $1\leq t-t_h<4$ and 
$5\leq t-t_h<8$.
There is an apparent, though slight, decreasing trend of the hardening 
rate with $N$ in the former interval, 
while the latter interval shows no identifiable trend.
We also give in Table \ref{tab_summary} the dimensionless hardening 
rate $H$ computed from the measured values of $da^{-1}/dt$ using
equation (\ref{defhard}); the stellar density and velocity dispersion
in that expression were evaluated by averaging inside a sphere of radius 
$r=0.04$ around the binary (for evaluating the velocity dispersion, 
we excluded the center $r\leq 0.01$ where stars are strongly 
perturbed by the binary).
We find that $H$ ranges from $\sim 6$ to $\sim 9$, consistent
with the moderately-hard binary results of 
three-body scattering experiments summarized above.
The lack of a noticeable $N$-dependence in the hardening rate is
consistent with the fact that the central stellar density and
velocity dispersion also do not vary substantially with $N$ 
(Figure \ref{fig_binary}d,e).

The orbital eccentricity of the BH binary evolves with time as well
(Figure \ref{fig_binary}c).
The eccentricity was evaluated from the binding energy and 
the angular momentum of the binary; before $t_h$, we corrected 
the BH masses by adding the mass in stars bound to each of them, 
$M_\bullet+M_*(r\leq a/2)$.
The initial orbital eccentricity of the galaxies is $\sim 0.75$
and this eccentricity is reflected in the relative orbit of the 
two BHs when they first fall to the center.
However the orbit rapidly circularizes due to the strong
density gradients in the cusp, 
and by $t=t_h$ the eccentricity is essentially zero.
From that point on, $e$ grows at an approximately constant rate, 
albeit with substantial fluctuations.
The fluctuations do not seem to be strongly correlated with $N$.
Final values of $e$ range from $\sim 0.15$ to $\sim 0.3$.
Table \ref{tab_summary} gives average values of the dimensionless
eccentricity growth rate $K$ obtained by fitting the integrated form 
of equation (\ref{eq_eccgrow}), assuming constant $K$,
to data for $t>11.6$.  In runs {\sf A2} and {\sf A4}, 
the growth of eccentricity is well approximated by the relation
\begin{equation}
\label{eq_eccfit}
e=K\ln\left(\frac{a_{ecc}}{a}\right)
\end{equation}
with $K=0.13$ and $a_{ecc}=0.001$.  (The three-body scattering results of 
\citet{qui96} predict a substantially lower rate, 
$K\approx 0.04$ for $e=0.3$ and $V_{bin}/\sigma_*=10.0$, although they 
allow for $K=0.13$ when $e\gtrsim0.4$ and $V_{bin}/\sigma_*\gtrsim30.0$.)
If the eccentricity continued growing at the 
rate predicted by equation (\ref{eq_eccfit}),
it would reach $e=0.5$ for $a\approx 2\times10^{-5}$, which is 
near the semimajor axis when the hardening rate due to gravity wave 
emission becomes larger than that due to stellar ejection
(\S\ref{sec_discuss}).  
Equation (\ref{eq_tgrav}) would then imply a gravitational 
radiation time scale $\sim 5$ times shorter 
than if the binary were circular.

Figure \ref{fig_binary}f shows the mass ejected by the binary
$M_{ej}$ as a function of time.
We monitored the mass ejection by counting stars with positive energies, 
or equivalently, with galactic escape velocities.  
This conservative criterion underestimates the number of stars ejected 
by a moderately hard binary ($V_{bin}/\sigma_*\lesssim 4$) because some 
of these stars are not energetic enough to escape the galaxy.
Ejected mass at the end of each run is close to $\mbin$, 
the total mass of the binary.
We fit $M_{ej}(t)$ to the integrated form of the mass ejection law, 
equation (\ref{eq_mejlaw}), to derive $J$ and $a_{ej}$. 
The fit was satisfactory for $t\gtrsim t_h+2$;
the fitted values of $J$ and $a_{ej}$ are given in Table
\ref{tab_summary}.
We find $J\approx 0.5$ which is consistent with the 
scattering experiments cited above.
At earlier times the mass ejection law with constant $J$
underestimates $M_{ej}/\mbin$.
As was the case for the hardening rate, the ejected mass does
not show a strong dependence on particle number in these simulations.

Readers should not compare the ejected mass in our simulations
with that in the simulations of \citet{quh97}; these authors measured greater
$M_{ej}$ for a smaller increase in $1/a$, but they also used a different, 
more liberal definition for $M_{ej}$. 
The primary focus of our study is to relate the dynamics of 
the BH binary's hardening to the {\it observable} responses of 
the stellar nucleus, such as the decrease in slope of the central cusp,
and the precise definition of $M_{ej}$ is not important for any of
the physical conclusions drawn here.

Although the supply of stars to the binary BH remains high
throughout our simulations, there is nevertheless a steady drop in the
stellar density as stars are ejected from the core.
This can be seen, indirectly, in the slight curvature of the
hardening rate plot (Figure \ref{fig_binary}b), and directly
in the change in density within a sphere of radius $0.04$
centered on the binary (Figure \ref{fig_binary}d).
More detailed information about the change in the stellar
mass distribution is given in Figure \ref{fig_profs}.
Most notable is the drop in central density between the first profile,
at $t=t_h$,  and the second that is offset in time by one $N$-body unit.
This is the same rapid drop in density that was discussed above (\S2),
associated with formation of the hard binary.
The amount of mass ejected during this short time interval is somewhat
greater with smaller $N$, which is also evident in Figure \ref{fig_binary}d.
We attribute this mild $N$-dependence to statistical fluctuations,
and also to a spurious effect associated with the wandering of the BH
binary:
since density profiles in Figure \ref{fig_profs} are centered
on the binary, densities at radii less than the wandering
radius may be artificially lowered.

There is no suggestion that a ``hole'' is forming around the BH binary;
apparently the supply of stars is great enough that the binary can eject
of order its own mass without driving the central density to a very
small value.
The central density profile, which is slightly steeper than $\rho\sim r^{-1}$ 
at $t=t_h$, 
becomes slightly flatter than $r^{-1}$ by the end of the integrations 
(Figure \ref{fig_profs}) and would presumably become ever flatter
if the integrations were continued to longer times.
The inner density profile remains well described as a power law at
all times.

\subsection{Brownian motion}

\begin{figure*}
\epsscale{1.8}
\plotone{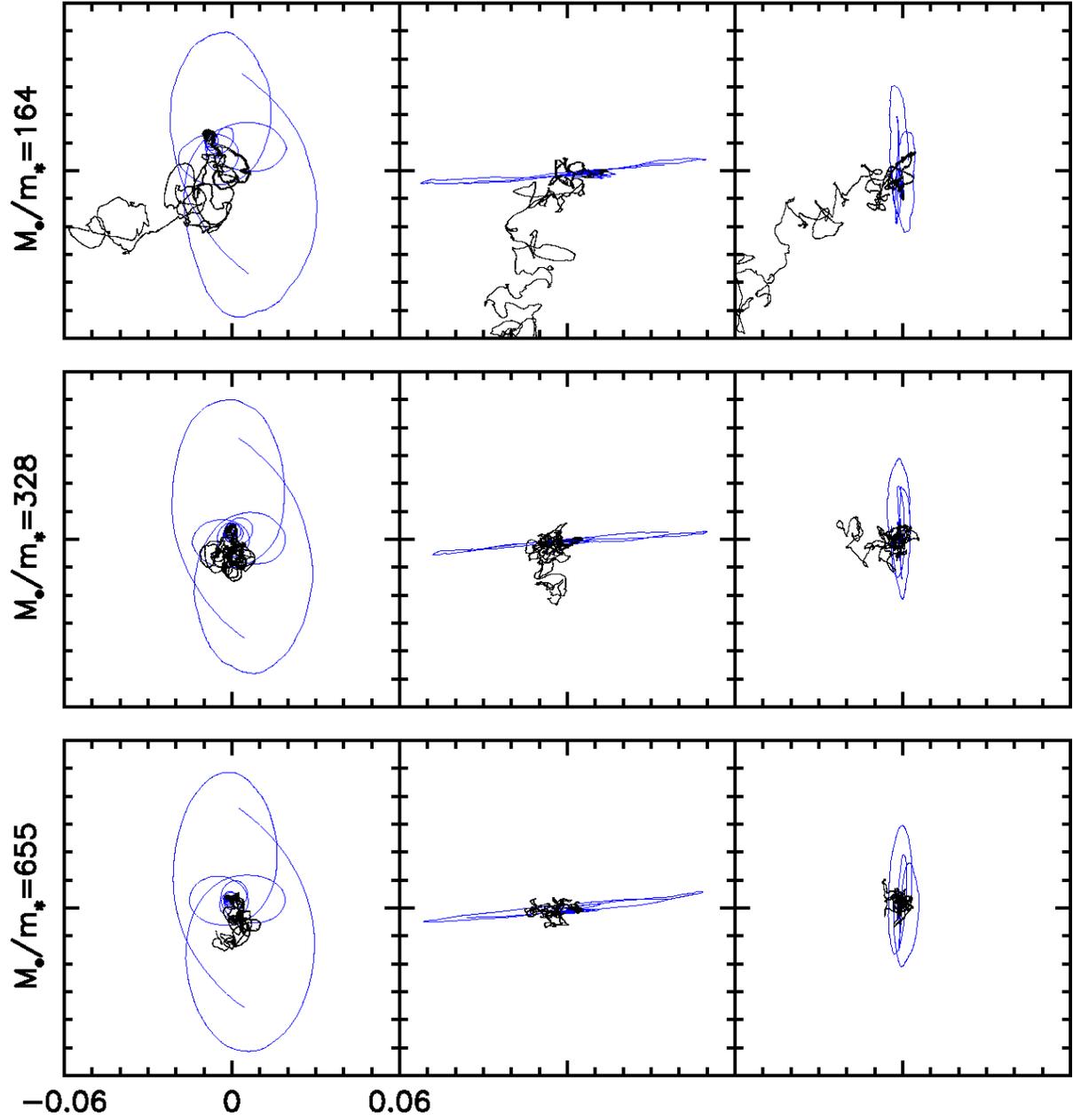}
\epsscale{1.0}
\caption{
Trajectories of the BHs before formation of a hard binary (blue curves) and 
after (black curves).  
The character of the motion changes suddenly to 
a random walk, characteristic of Brownian motion, after the binary forms.
The amplitude of the Brownian motion decreases with increasing
BH mass (as indicated on the left).\label{fig_wander}}
\end{figure*}

\begin{figure}
\plotone{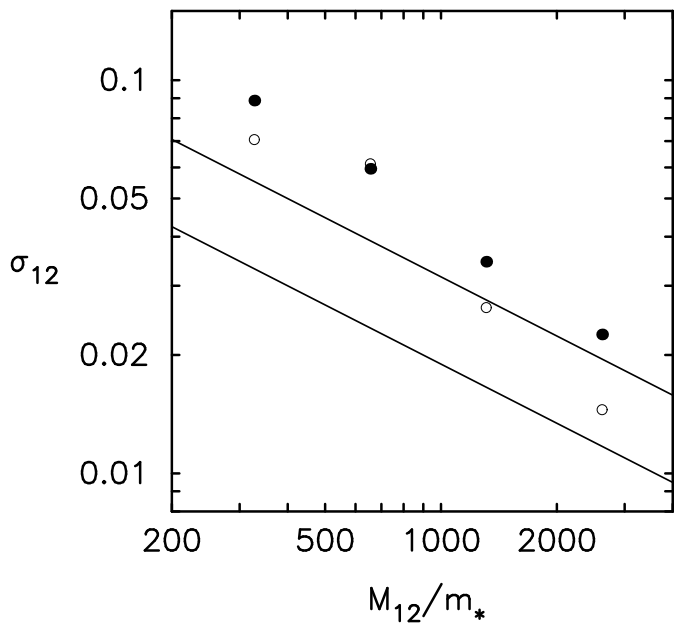}
\caption{
1D rms center-of-mass velocity of the BH binary's Brownian motion, 
in the merger plane (filled circles) and perpendicular to the merger 
plane (empty circles).
Equipartition of energy implies $\sigma_{12}=(M_{12}/m_*)^{-1/2}\sigma_*$ 
which depends on the region over which we average $\sigma_*$.  
Solid lines show the band of values consistent with equipartion between 
$\sigma_*=0.6$ ($r<0.2)$ and $\sigma_*=1.0$ ($r<0.01$), 
the binary's radius of influence.\label{fig_equipart}}
\end{figure}

Figure \ref{fig_wander} shows the trajectory of the two BHs 
at high spatial resolution in our simulations.
The Brownian motion is apparent 
as a sudden change in the character of the BH orbits
at $t\approx t_h$.
Prior to this time the trajectories are smooth and symmetric, 
reflecting the dynamical-friction induced coalescence of the two cusps.
However starting at $t\approx t_h$ the motion becomes more chaotic, 
resembling a random walk.
Figure \ref{fig_wander} shows clearly that the amplitude of the
random motion is a decreasing function of $N$, as expected
from equipartition arguments (equation \ref{equipart}).
We quantified the $N$-dependence by computing 
$\sigma_{12}\equiv\sqrt{\langle v^2\rangle/3}$ 
for the binary's center of mass.
This quantity exhibits almost no evolution with time 
(Table \ref{tab_summary}); 
Figure \ref{fig_equipart} shows averages for $t > t_h$.
The equipartition relation $\sigma_{12}\propto \mbin^{-1/2}$
is approximately satisfied.
Of interest is the amplitude of $\sigma_{12}$ which is expected
to be slightly larger for a binary BH than for a single BH
\citep{mer01}.
We were unable to check this prediction by a direct comparison
between our two- and single-BH runs since the latter were not
extended long enough that an accurate characterization of the
BH's Brownian motion could be obtained.
Instead, we compared $\sigma_{12}$ with the velocity dispersion
predicted by equation (\ref{equipart}), 
$\sigma_{12}=(m_*/\mbin)^{1/2}\sigma_*$.
This requires a choice about how to evaluate $\sigma_*$, which
depends weakly on radius.
In Figure \ref{fig_equipart} we plot a range of predictions for
$\sigma_{12}$ based on measured values for $\sigma_*$ within
$r=0.2$ ($\sigma_*\approx 0.6$) and $r=0.01$ ($\sigma_*\approx 1.0$).
The rms velocity of the binary appears to be slightly greater than
expected for a point mass, as predicted by \citet{mer01}.

Brownian motion of a massive binary was described in a few earlier studies.
\citet{quh97}, in a series of $N$-body simulations,
noticed a wandering of their BH binary with an amplitude
$5-10$ times greater than expected on the basis of an equation like
(\ref{defwander});
they attributed the discrepancy to inelastic scattering of ejected stars.
\citet{mak97} carried out $N$-body simulations similar to those of
\citet{quh97} but using a more conservative,
direct-summation code and no mass spectrum for the field stars.
Makino's Figure 7 shows a wandering amplitude that scales as $\sim N^{-1/2}$,
and the rms velocity of the binary appears to be
comparable to that expected for a point mass.
Makino's results seem consistent with those obtained here
and with the predictions of \citet{mer01}: a massive binary 
at the center of a dense cusp should exhibit only slightly greater
Brownian motion than a point particle of the same total mass.
It is not clear why \citet{quh97} found a much greater
amplitude for the Brownian motion in their simulations;
some possible reasons for the discrepancy are discussed in 
\citet{mer01}.

\begin{figure}[t]
\plotone{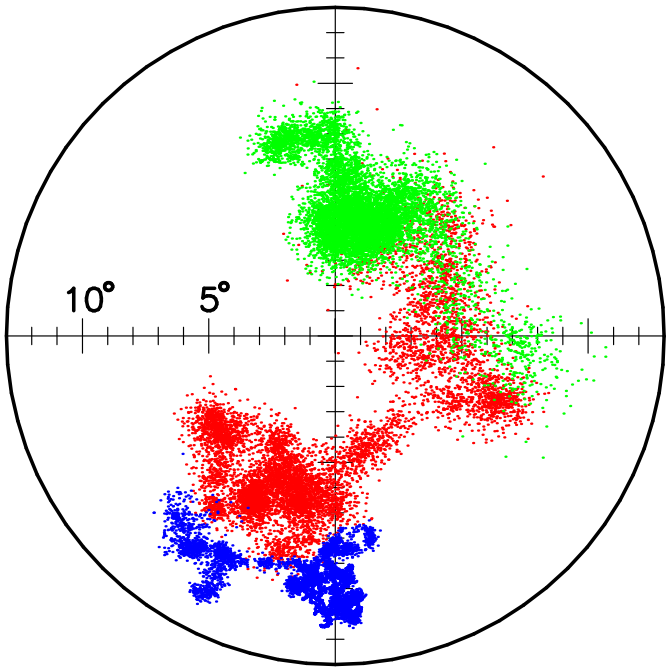}
\caption{
Angular inclination of the BH binary's axis of rotation. 
Red dots: run {\sf A8} ($M_\bullet/m_*=164$);
green dots: run {\sf A4} ($M_\bullet/m_*=328$); 
blue dots: run {\sf A2} ($M_\bullet/m_*=655$).   
Binary's angular momentum unit vector in polar coordinates 
$\hat n(\theta,\phi)$ is shown as a dot at position 
$(\theta\cos\phi,\theta\sin\phi)$.  
Center of the plot corresponds to rotation in the merger plane.  
Inclination of the binary undergoes a random walk with an amplitude 
that decreases with increasing $M_\bullet/m_*$.
\label{fig_dir}}
\end{figure}

The orientation of the binary is also affected by encounters.
Figure \ref{fig_dir} shows the direction of the binary's
angular momentum vector ${\bf \hat n(\theta,\phi)}$ where each dot represents
the angular tilt of ${\bf \hat n}$ from the merger axis.
Points shown correspond to $t>t_h$;  
before the binary is hard, its orientation changes at much higher 
rate, albeit with similar amplitude, due to transient bulk 
perturbations occurring during the galactic merger.  
After $t_h$, the bulk torques are negligible, but 
the orientation still changes due to torques imparted by 
elastic and inelastic encounters with stars.
It is evident in 
Figure \ref{fig_dir} that the net effect of stellar ejections is a
random walk in the tilt-space $(\theta,\phi)$ and that the rate of 
orientation-changing decreases both with hardness and with $N$.  
We observed a maximum tilt of $\theta\approx 12^\circ$ degrees 
from the merger axis.

\subsection{$N$-dependence of the evolution}

As discussed above, we do not observe an appreciable dependence
of the binary hardening rate on $N$ in our simulations.
This is reasonable since the expected hardening rate (equation \ref{defhard})
depends only the mean density $\rho$ and velocity dispersion $\sigma_*$ 
of the field star distribution, not on the masses of field stars.

Some earlier studies noticed a more appreciable $N$- dependence.
\citet{quh97} found that the decay rate dropped with
$N$ until $N\approx 10^5$, then seemed to level off at $N=2\times 10^5$.
\citet{mak97} described the $N$-dependence of the decay rate as
weaker than $N^{-1}$ but gave no further details.
In both of these studies, 
the $N$-dependence of the hardening rate was attributed
to the Brownian motion, since larger $N$ implies a reduced amplitude
of wandering and hence a smaller
pool of stars that can interact closely with the binary.
Quinlan \& Hernquist showed that the binary's hardening rate dropped 
rapidly to zero if the binary was artificially fixed in space.
This result is also implicit in the study of \citet{zie00} who
calculated the rate of depletion of stars around a massive binary
that was fixed in space.

Why do we fail to see any clear $N$-dependence of the hardening rate in our
simulations -- given that the Brownian motion {\it does} vary in
the expected way with $N$ (Figure \ref{fig_equipart})?
The main reason, we believe, is our very different initial conditions,
which guarantee a larger supply of stars to the binary than in earlier studies.
We noted above that the models
of \citet{mak97} and \citet{quh97} had much lower
central densities than ours at the time of formation of the hard binary.
The supply of stars was correspondingly smaller,
implying a more rapid depletion of the ``loss cone;'' 
once this occurs, the supply of stars is essentially cut off and any
further decay depends on $N$-dependent processes such as loss-cone
refilling and Brownian motion.
The origin of the much lower initial density in the 
\citet{quh97} simulations may be seen in their Figure 1c, 
the evolution of the Lagrangian radii 
for a run where the BHs have a combined mass that is 1\% of
the galaxy mass, comparable to the value in our simulations.
The stellar mass within a sphere of radius $0.01$ drops
from $\sim 1\%$ initially to $\sim 0.05\%$ by the time of formation
of the hard binary.
In our simulations, the drop is from $\sim 1\%$ to only $\sim 0.3\%$
over a comparable interval of time (Figure \ref{fig_lagrange}).
The reason for the much greater density drop 
in the Quinlan \& Hernquist simulations was their 
choice of initial conditions:
the BHs were initially placed far {\it outside} 
the central cusp and fell in.
Makino's (1997) models had large cores from the start.

\begin{figure}[t]
\plotone{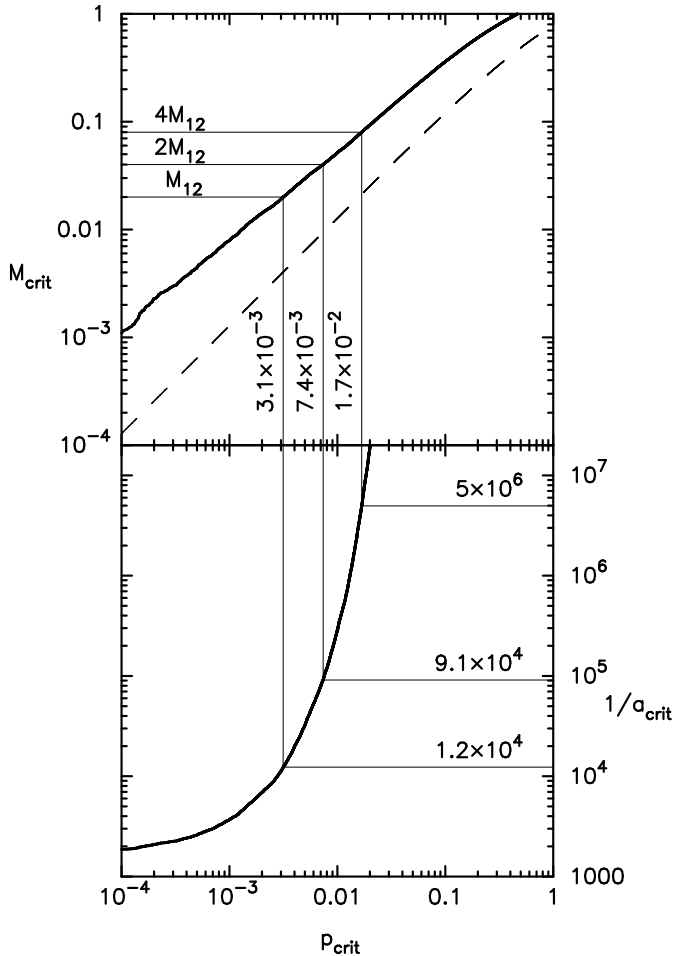}
\caption{
Relation between the ejected mass and the shrinking of the BH binary.
Upper panel is the mass in stars with pericenter distances smaller 
than a given value $p_{crit}$ as function of $p_{crit}$
(solid curve).  
For comparison, we show the mass in 
stars inside radius $r=p_{crit}$ (dashed curve).  
$M_{crit}$ is the mass accessible to the BH binary provided that 
it can visit every point within a distance $p_{crit}$ 
from the center of the galaxy.  
Lower panel is 
the maximum inverse semimajor axis $1/a_{crit}$ at 
which the binary will stall if it can interact only with 
stars having pericenter distances less than $p_{crit}$ (solid curve).
\label{fig_peri}}  
\end{figure}

The ``supply'' of stars in our models can be defined, very approximately, 
as the number of stars (at $t=t_h$, say) with pericenters less than some 
critical value $p_{crit}\approx a$.
This definition ignores changes in the stellar density profile that result
from the changing potential, as well as any back-reaction of the 
binary's motion on the stellar distribution.
We also ignore any dependence of $p_{crit}$ on stellar velocity,
even though low-velocity stars have a larger cross section
for interaction with the binary than high-velocity stars.
Figure \ref{fig_peri} shows $M_{crit}$, the mass in stars
with pericenters below $p_{crit}$, as a function of $p_{crit}$. 
This was computed by counting orbits penetrating, or entirely contained within,
 the sphere of radius equal to
$p_{crit}$ in an $N=2^{18}$ particle realization of our 
initial model (equation \ref{eq_jaffe}) of pre-merger galaxies.  
For $p_{crit}\lesssim 0.1$, the function is approximately a power law
\begin{equation}
\label{eq_mcrit}
\frac{M_{crit}}{M} \approx 1.8\times \left(\frac{p_{crit}}{r_0}\right)^{0.84}
\end{equation}
where $M$ is the mass of the galaxy and $r_0$ is the half-mass radius.  
The figure also shows $1/a_{crit}$, the inverse semimajor axis
attainable for an ejected mass of $M_{crit}$.
This was computed from the relation
\beq
{a_{crit}\over a_{ej}}  =  e^{M_{crit}/J\mbin} 
\label{eq_acrit}
\eeq
using the fitted values of $J$ and $a_{ej}$ (Table \ref{tab_summary}).  
Taken together,  equations (\ref{eq_mcrit}) and (\ref{eq_acrit}) 
relate the size of the region accessible to the binary ($p_{crit}$)
to the minimum semimajor axis that the binary can reach by ejecting all 
stars visiting the region ($a_{crit}$).

In order for the binary to have access to sufficient mass allowing
it to decay to $a^{-1}\approx 10^4$, roughly the final value in our
simulations, Figure \ref{fig_peri} suggests that
it must eject all stars with pericenters lying within $\sim 0.003$.
This is comfortably {\it smaller} than the wandering radius
even in our largest-$N$ simulation 
($r_w\approx 0.01$; Table \ref{tab_summary}).
We believe that this explains why none of our binaries have managed to 
``deplete the loss cone'' and stall -- their supply of stars was
never depleted, at least over the interval of time represented by 
the simulations.
Combining Figures \ref{fig_equipart} and \ref{fig_peri},
we estimate that $N$ would have to be increased to $\sim 5\times 10^5$
per galaxy in order for the Brownian motion to be small enough that
the supply of stars would be exhausted by the time that $1/a=10^4$.

Given that the ``loss cone'' is never depleted in our simulations,
how great is the effect of two-body relaxation on the decay rate of the
binary?
We estimated, using standard expressions like equation (\ref{losscone}),
that scattering of stars into the loss cone probably does make
a significant contribution to the decay rate of the binary in
our lowest-$N$ run.
For the larger $N$ runs it is probably of negligible importance.
This conclusion would presumably change if the loss cone ever became
fully depopulated.

What would happen in this case -- if the supply of stars to the binary
were depleted, either by continuing the integrations to much later times,
or by using a larger $N$ and thereby reducing the amplitude of the binary's
Brownian motion?
Our guess is that the decay would in fact stall, producing a 
BH binary whose separation remained nearly constant for extended
periods of time.
The same would presumably also result from initial conditions with
much lower central density, for instance, the merger of two giant 
elliptical galaxies.
We return to the question of the persistence of BH binaries in
\S\ref{sec_discuss}.

We caution that our simple picture, 
of a BH binary wandering against a fixed distribution of stars,
is not completely correct.
In fact the binary ``carries'' the cusp with it to a certain extent.
This is clear from the density profiles plot, Figure \ref{fig_profs}:
the cusp continues as a power law to radii much smaller than the wandering
radius.
It follows that the wandering of the binary is a complex process
involving time-dependent changes in the stellar distribution,
and these changes probably affect to some extent the supply
of stars to the binary in a way not reproduced in our simple
analysis.
Brownian motion of a massive object in a density cusp would be
a fruitful topic for further study.

\section{Kinematics}
\label{sec_kin}

Just as the BH binary affects the stellar density profile at distances as 
large as the break radius, so we expect the presence of BHs to
shape the remnant's kinematical properties well beyond the binary's
gravitational radius of influence.  
Here we present simulated ``observations'' of our largest-$N$
merger remnants, from runs {\sf A2} and {\sf B2},
in zero dimensions (circular apertures), one dimension (slit)
and two dimensions (integral field).
The purpose is to generate predictions that can be tested with
the current generation of high-spatial-resolution spectrographs
such as {\sf STIS}, {\sf OASIS} and {\sf SAURON}.
Following this, we present several revealing
views of the remnant in ways that are not directly accessible to 
astronomical observation.

We ``observed'' our galaxies in two steps
(Appendix \ref{ap_losvd}).
Starting from the stellar velocities projected into an edge-on view of the
galaxy, we recovered the line-of-sight velocity distributions (LOSVDs)
non-parametrically via maximum penalized likelihood (MPL)
\citep{mer97}.  
The MPL estimate $\hat N(V)$ of an LOSVD is computed on a
grid in velocity such that it maximizes the log-likelihood of the
distribution of line-of-sight projected stellar velocities inside an
aperture, subject to a penalty function that measures the lack of smoothness of
$\hat N(V)$.  
Once $\hat N(V)$ was obtained, we expanded it into its Gauss-Hermite 
(GH) moments defined by \citet{ger93} according to the prescription of 
\citet{mfx93}.  
Of particular interest are the four parameters in the GH
expansion ($V_0$, $\sigma_0$, $h_3$, $h_4$) that quantify, respectively,
the mean velocity and velocity dispersion of the Gaussian prefactor, 
and the odd and even first-order departures from a Gaussian distribution.  
(Henceforth in this section, the terms ``mean velocity'' and 
``velocity dispersion'' are used in this restricted sense.) 
These parameters, however, are insensitive to the power-law wings 
expected in LOSVDs in the vicinity of a BH \citep{vdm94}
and hence it is important to consider the full LOSVD.

To increase the resolution inside each circular aperture of diameter $D$,
we superposed one hundred snapshots of the galaxy that were sampled over
$1$ $N$-body unit in time, corresponding to $\sim 10$ crossing times at
a radius $r\sim 0.1$ from the center.  
Averaging over such a wide time interval
ensures that stars in the aperture are sampled at random orbital phases.
In this procedure each dataset was shifted in space so that the BH binary
lay at the center.  
The centering could cause spurious smoothing on scales
smaller than the radius of the Brownian motion of the binary (cf.
\S\ref{sec_hard}) which however amounts to not more than $r_w\approx 0.0084$ in
the run {\sf A2} that was used for this purpose.  When superposing
datasets, the stellar velocities were left in their original frame.  
Orbits closely bound to the BH binary may follow the binary on its random
Brownian trajectory, thereby incurring a net drift in the velocity.  This
drift, however, scales as $(m_*/M_{12})^{1/2}\sigma_* \approx 0.05$ and
can be ignored at this stage.

\begin{figure}[t]
\plotone{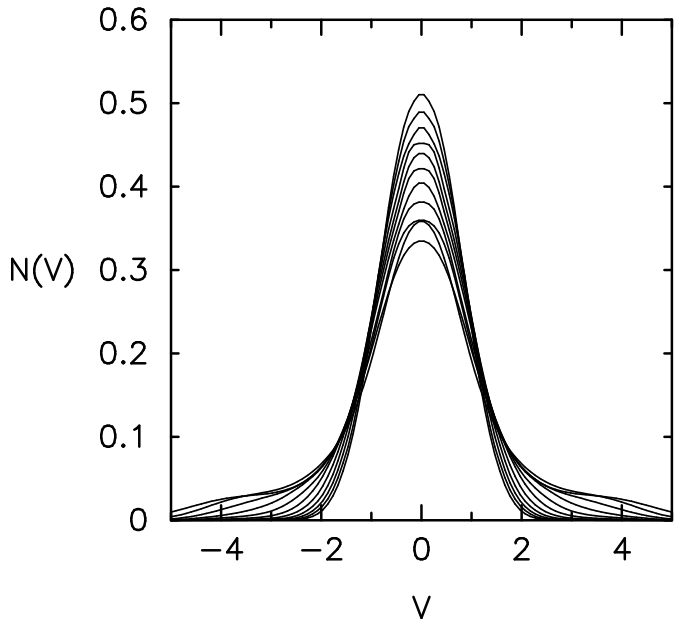}
\caption{
Line-of-sight velocity distributions (LOSVDs) of the 
merger remnant in the run {\sf A2} 
as a function of aperture diameter $D$.
The aperture is circular and centered on the BH binary; 
its diameter varies from $D=0.002$ (LOSVD with
 the shortest peak and the broadest wings) to 
$D=0.2$ (LOSVD with the tallest peak and the narrowest wings).
\label{fig_losvds}}
\end{figure} 

We anticipate the discussion in \S\ref{sec_discuss}.1 by quoting
typical values for the physical scale of our models.
Scaling to a dwarf elliptical galaxy like M32
gives a factor ${\cal L}\approx 50$ pc for converting
model dimensions into physical lengths.
In the case of a giant elliptical galaxy like M87, this factor is
$\sim 3$ kpc.
Note that the radius of gravitational influence $r_{gr}$ of the BHs is
$\sim 0.02$, independent of the scaling.

\begin{figure}[t]
\plotone{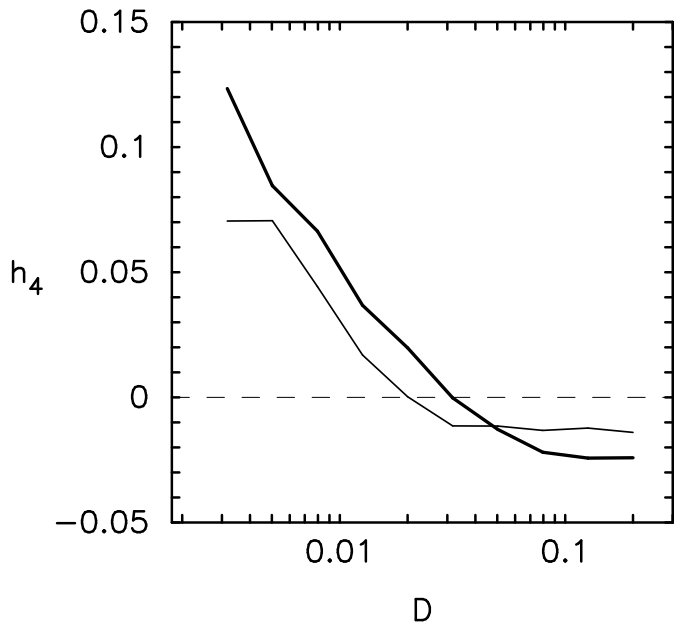}
\caption{
Fourth Gauss-Hermite moment $h_4$ as a function of aperture diameter $D$.
The aperture is centered on the BH binary in the run {\sf A2} (thick line) and on the single BH in the run {\sf B2} (thin line).
\label{fig_h4}}
\end{figure}

In Figure \ref{fig_losvds} we show a family of LOSVDs from the run {\sf
A2} for a range of diameters $D$ of a circular aperture centered on the BH
binary.  
The LOSVD that is most nearly Gaussian is seen inside an aperture
with diameter about twice the BH binary's radius of influence, $D\sim
2r_{gr}$.  
The LOSVD seen in the smallest diameter aperture $D=0.002$ has
the shallowest peak and broadest wings, 
while the one seen in the largest diameter aperture $D=0.2$ has 
the steepest peak and almost non-existent wings.  
These differences are reflected in the values of $h_4$ shown in Figure 
\ref{fig_h4}.  
In general, positive $h_4$ indicates that the LOSVD is sharp 
(or ``triangular'') at the top and decays more mildly
on the sides;
negative $h_4$ indicates that the LOSVD is broad (or ``boxy'') at the top
and steep on the sides, reflecting a sharp maximum velocity cutoff.

\begin{figure}[t]
\plotone{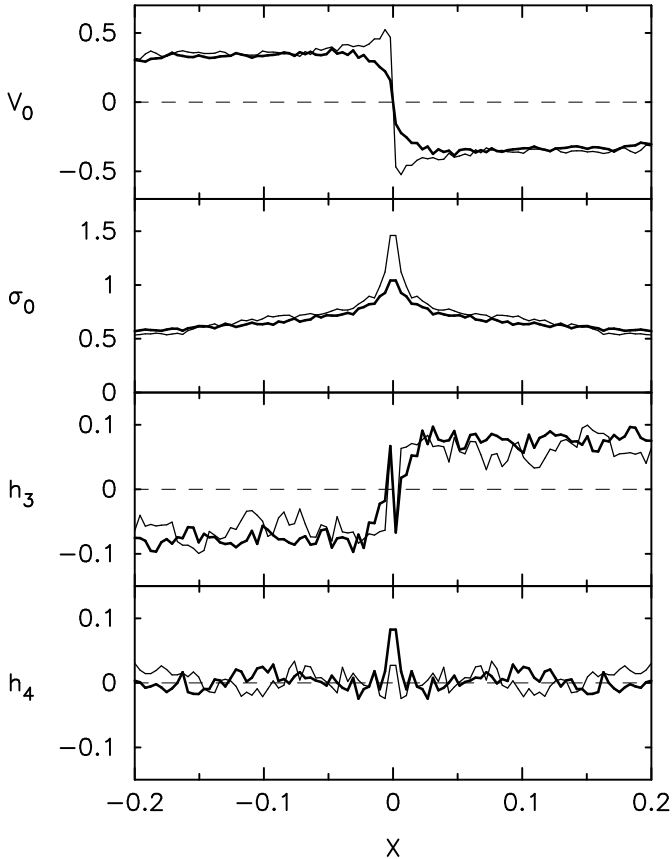}
\caption{
Slit kinematics of the run {\sf A2} (BH binary; thick line) and {\sf B2} 
(single BH; thin line).  
Slit is positioned along the major axis and centered on the BH.  
The aperture diameter (slit width) varies from $D=0.02$ at $X=\pm0.2$ to $D=0.004$ at the center.  
Parameters $V_0$, $\sigma_0$, $h_3$ and $h_4$ in the Gauss-Hermite 
expansions of the LOSVDs were calculated as described in the text.
\label{fig_slit}}
\end{figure}

We observe that $h_4$ decreases from $h_4\approx 0.12$ for $D=0.003$ to
$h_4\approx -0.024$ for $D=0.2$ passing through zero for $D\approx 0.03$.  
A similar trend was derived by \citet{vdm94} under the assumptions of isotropy and
spherical symmetry for a model of M87 with a $5\times10^9\msun$ BH; the
size of the aperture influences only the overall normalization, and not
the velocity dependence, of the LOSVD in the large-velocity limit.  
For comparison, in Figure \ref{fig_h4} we also plot $h_4$ from the run {\sf
B2} where two BH were combined into one at $t=t_h$.  The absolute
amplitude of $h_4$ around the single BH is smaller than around the binary,
which is counterintuitive in view of the even wider wings (not shown here)
that we found for small apertures in the run with one BH.  The
Gauss-Hermite moments, however, are sensitive to the velocity profile in
the range $V\lesssim 2\sigma_0$ and indifferent to the
high-velocity behavior.  
It is therefore possible that an LOSVD is both
``boxy'' at low velocities ($h_4 \lesssim 0$) and ``wingy''
in high velocities (which may or may not imply a positive $h_4$).

In Figure \ref{fig_slit} we plot the major axis slit kinematics of the merger
remnant in the run {\sf A2} (BH binary) and {\sf B2} (single BH).
To increase the resolution, we added together six views of the major axis,
rotated by angles $k\pi/3$, $k=0,1,2,3,4,5$, around the minor 
axis, from the original line of sight ($k=0$).
This yields $V_0$ and $h_3$ that are odd and $\sigma$ and
$h_4$ that are even under reflection $X\rightarrow-X$.  
We first established  the
position-dependent maximum aperture diameters ($D=0.004$ at the center and
$D=0.02$ at the ends) and then narrowed the apertures with the requirement
that the number of stars inside each aperture remain larger than a fixed 
amount ($4096$), 
ensuring homogeneous statistics across the slit and uncorrelated
sampling ($D<\Delta X$) at the center.

\begin{figure*}
\epsscale{1.6}
\plotone{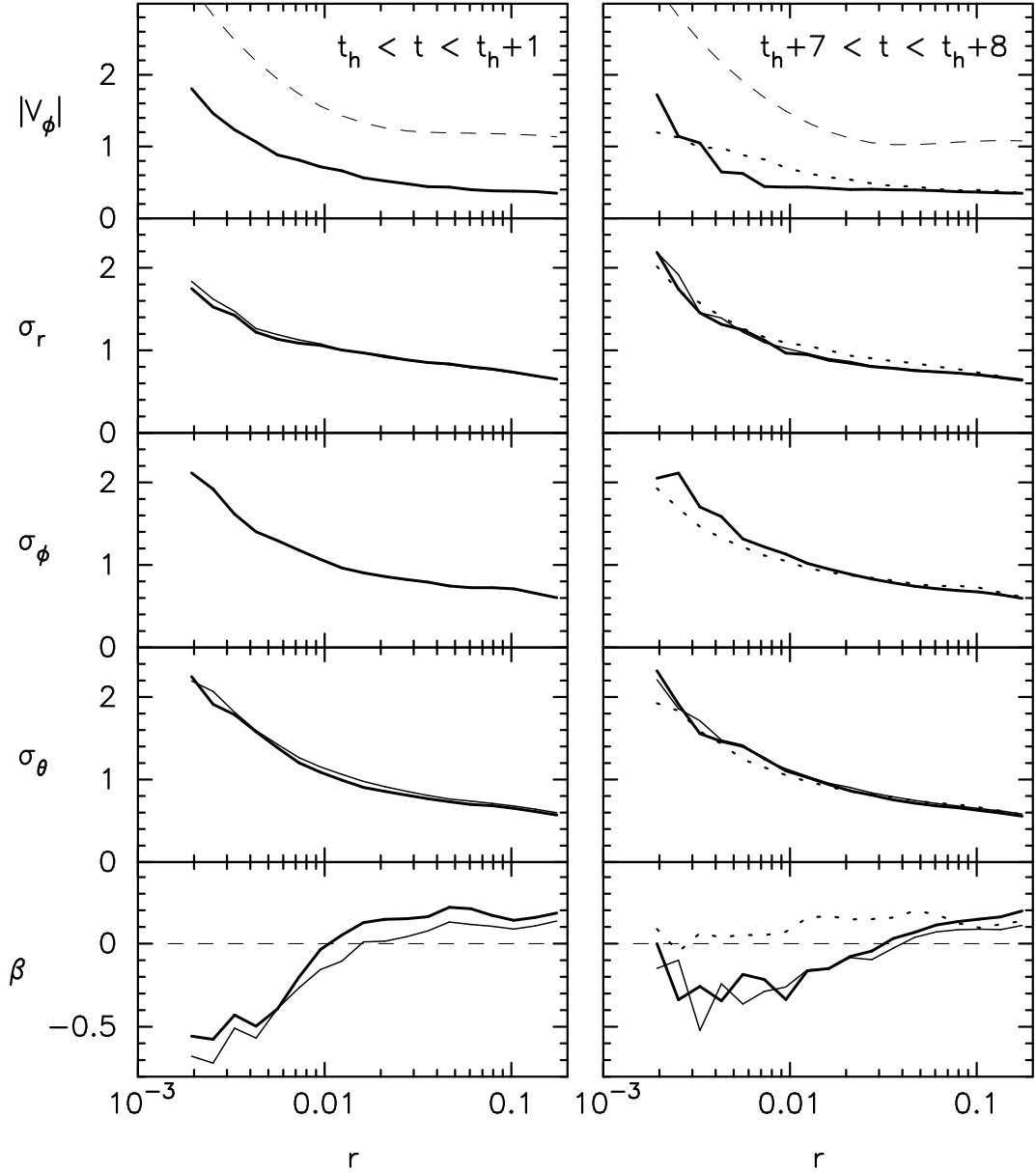}
\epsscale{1.0}
\caption{
Internal kinematics along the major axis (thick curve) and minor axis 
(thin curve) immediately after the formation of a hard binary (left column) 
and at the end of simulation (right column).  
Solid curves: run {\sf A2} with a BH binary.  
Dotted curves: run {\sf B2} with single BH (major axis only).
The major and minor axis moments were computed by direct averaging inside the circular wedges $|\theta-90^\circ|<30^\circ$ and $|\theta-90^\circ|>30^\circ$, respectively.  
First row: rotational velocity $v_\phi$; 
thin dashed curve is the circular velocity $v_c\equiv\sqrt{G(M_{12}+M(r))/r}$. 
Second to fourth rows: moments of the rms velocity dispersion ($\sigma_r$, $\sigma_\phi$, $\sigma_\theta$).  
Fifth row: anisotropy parameter $\beta\equiv 1-\sigma_t^2/\sigma_r^2$ where $\sigma_t^2\equiv(\sigma_\phi^2+\sigma_\theta^2)/2$.  
The tangentially-anisotropic central region ($\beta<0$) grows with time, from $r\approx0.01$ at $t=t_h$ to $r\approx0.03$ at $t=t_h+7$.  
The simulation with a single BH exhibits no appreciable anisotropy.
\label{fig_moments}}
\end{figure*}

We also computed spatial (unprojected) kinematical quantities,
in particular, the rotational velocity $v_\phi$ and three diagonal 
moments of the velocity dispersion tensor 
$\sigma_r,\sigma_\phi,\sigma_\theta$.
This was done by averaging inside circular wedges of opening angle 
$|\theta-90^\circ|<30^\circ$ (major axis) and
$|\theta-90^\circ|>30^\circ$ (minor axis).  
Figure \ref{fig_moments} plots spatial kinematical properties of 
the galaxy immediately following the merger (left column) 
and of the final galaxy (right column) along the major and minor axes.
We also provide kinematical moments of the merger remnant where 
the two BHs were coalesced into one at $t=t_h$ (dotted line).  
We will argue below (\S\ref{sec_discuss}.4) that this model might
be a good representation kinematically of a ``power-law'' elliptical
galaxy.

All major elements of the slit kinematics that we observe in the simulations
are also found in the kinematical profiles of real galaxies.  

1. {\it Rotation curve}.
Our models exhibit flat or slightly falling rotation curves
at the outer radii, $|X|\gtrsim 0.05$, with $V_0/\sigma_0\approx 0.5$.  
At these radii, the runs with one and two BHs are essentially identical
kinematically.  
At inner radii, $|X|\lesssim 0.05$, the two rotation curves differ 
substantially.  
The rotation curve near the single BH rises and
appears to diverge near the center indicating a nearly-Keplerian rotation
pattern dominated by the BH's potential.  
Near the binary BH, however, the rotation curve drops, 
with a $\sim50\%$ smaller $V_0$ at the knee marking the drop than in
the case with a single BH.  
Our choice of the terms ``inner'' and ``outer'' is not arbitrary; 
in fact, transition between the two regions coincides with the
break radius defined above (\S\ref{sec_merge}).
We therefore suggest that the same physical mechanism
responsible for the shallowing of the nuclear density cusp also manages to
somehow attenuate the circumnuclear rotation.

We propose a mechanism closely related to mass ejection by a BH
binary (cf. \S\ref{sec_hard}) that leads to precisely this effect.  
When two stars pass near the binary in opposite directions, the
star on a co-rotating (prograde) orbit is more likely to be captured
by one of the BHs since it can interact with the BH over a
larger orbital phase than the counter-rotating (retrograde) star.  
As co-rotating stars are preferentially ejected from the nucleus, 
there will be an increase in the relative number of
counter-rotating stars, thereby attenuating the net rotation.  
In our simulations this effect is too small
to reverse the direction of rotation in the nucleus but it plausibly
explains the $50\%$ difference between the runs with one and two BHs.  
This interpretation implies a concrete physical prediction:
rotation curves in galaxies with shallow central density cusps (or ``core''
galaxies, cf. \S\ref{sec_discuss}) should turn over near the break radius and
exhibit systematically lower rotation to within several dynamical radii of
the BH than those in galaxies with steep density cusps (or ``power law''
galaxies).  

The rotation curve of our single-BH (``power-law'') model
looks similar to that of M32 as observed with {\sf STIS} 
on HST \citep{jos00}.

2. {\it Velocity dispersions.}
The central velocity dispersion exhibits a sudden upturn at a 
distance $\sim r_{gr}\approx 0.02$ from the BHs.
The spike is more pronounced in the run with one BH
($\sigma_{0,max}=1.46$) than in the run with two ($\sigma_{0,max}=1.04$).  
This difference is reduced when the ``corrected'' velocity dispersion
$\sigma\equiv\sigma_0(1+\sqrt{6} h_4)$ (not shown) is used;
$\sigma$ is a closer approximation than $\sigma_0$ to the true rms
velocity \citep{mfx93}.  
In the run with one BH, $\sigma_{max}=1.56$, while with two BHs 
$\sigma_{max}=1.26$.  
Nevertheless, in both the uncorrected and corrected velocity dispersions, 
we note a systematically lower value around the binary BH
compared with the single BH out to a radius of $\sim 0.1$.
Lower dispersions around the BH binary may at first sight strike the reader 
as unexpected, given that the binary injects energy into the system 
as it hardens.  
But the loss cone is populated largely by radial orbits;
as the binary captures, ejects, and removes radial orbits from the
nucleus, the radial dispersion $\sigma_r$ drops while the tangential
dispersion $\sigma_t\equiv\sqrt{\sigma_\phi^2+\sigma_\theta^2}$ remains
constant, resulting in a decrease in the average dispersion
$\sigma=\sqrt{\sigma_r^2/3+2\sigma_t^2/3}$.

\begin{figure*}[t]
\epsscale{2.0}
\plotone{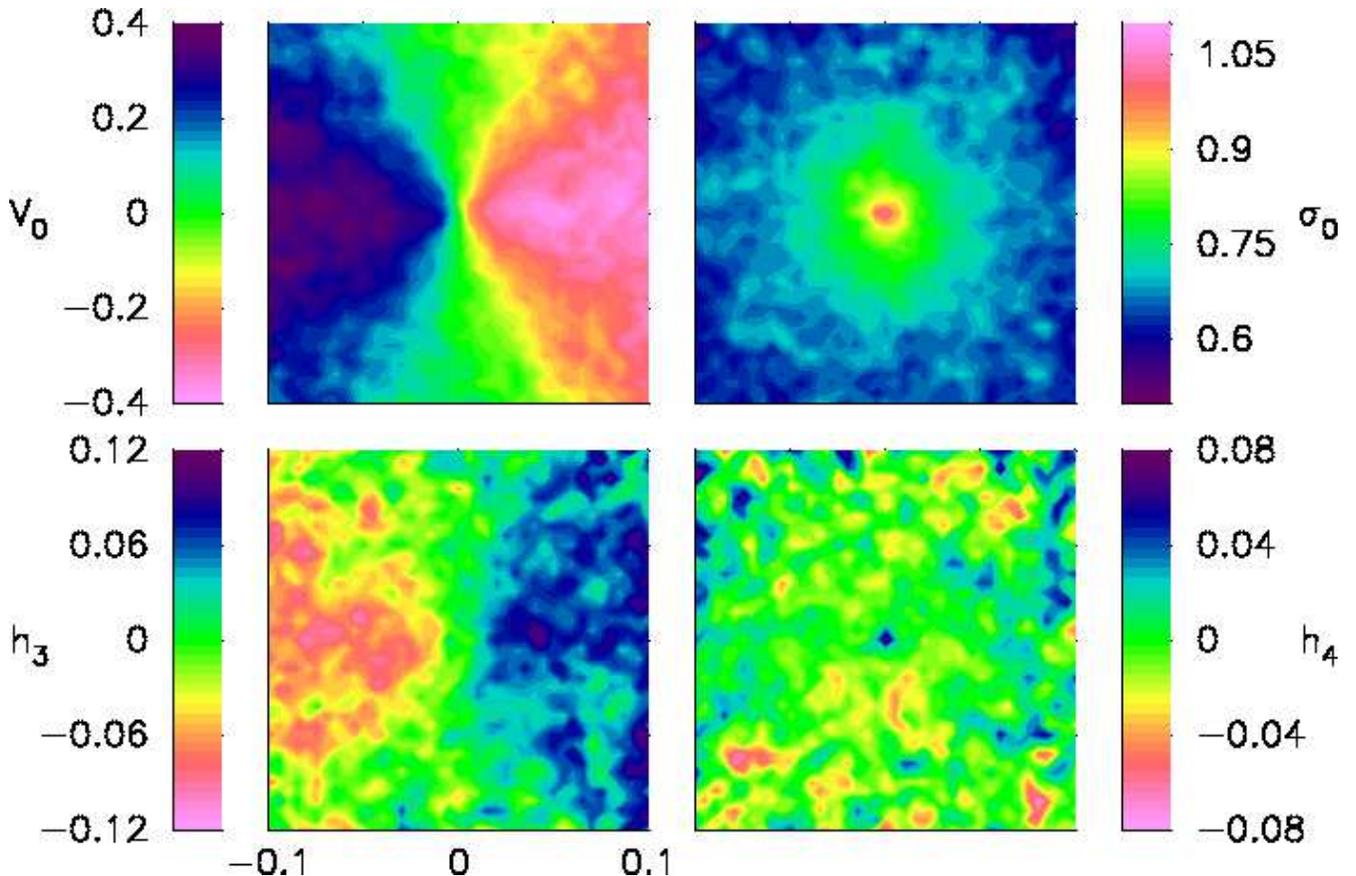}
\epsscale{1.0}
\caption{
Two-dimensional kinematical maps of the merger remnant from run {\sf A2}.  
The line of sight is parallel to the plane of the merger; 
the latter projects into the horizontal, or major, axis of the image.  
Gauss-Hermite parameters ($V_0,\sigma_0,h_3,h_4$) were recovered as
described in the text (\S\ref{sec_kin}).
To increase the resolution inside each aperture, 
100 snapshots from $18.1\leq t\leq 19.1$ were superposed; 
this interval amounts to $\gtrsim 10$ crossing times at radii 
$r\lesssim 0.1$ from the center.\label{fig_gh}} 
\end{figure*}

We in fact observe tangentially-anisotropic motions 
out to radii $r\lesssim 0.03$, as shown in Figure \ref{fig_moments}. 
In the bottom right panel we show the 
variation of the anisotropy parameter $\beta$ with radius in
the final models; $\beta$ is defined as
 $\beta\equiv 1-\sigma_t^2/\sigma_r^2$.  
In the case of a single BH, $0\leq\beta \leq 0.2$ at all radii,
which we believe is consistent with $\beta\sim 0$.  
In contrast, in the binary BH model, 
$-0.4\leq\beta\leq 0$ for $r<0.03$ and $0\leq\beta\leq
0.2$ for $r>0.03$.  
The anisotropy that we measure outside the binary's radius
of influence ($r_{gr}\approx 0.01$) is consistent with that from the 
equivalent run of \citet{quh97}.  
These authors detected $\beta\approx -1.0$ but only at radii
so close to the center that the calculation of $\beta$ may
depend on the resolution effects and the binary's Brownian motion.  
\citet{zie00}, whose binary was fixed in space, found even stronger
central anisotropy.  
We reiterate the prediction of \citet{quh97} that
weak density cusps formed by the action of BH binaries are tangentially
anisotropic.  
We however disagree with the characterization of the
anisotropy as ``strong;'' in fact it is mild
$(\beta\lesssim 0.5$) on scales that can be resolved observationally.  
On smaller scales, we warn against the possibility of contamination of any
inferred anisotropy by stars that have recently been ejected or that
are interacting strongly with the binary.

3. {\it Third Gauss-Hermite moments.}
$h_3$ is constant and has a sign opposite to that of 
the velocity parameter $V_0$.  
If $h_3$ and $V_0$ have the
same sign, the prograde wing of the LOSVD is wider and the retrograde wing
is steeper; if the signs are opposite, the reverse is true.  Figure
\ref{fig_slit} shows $h_3$ in the runs with one and two BHs; in the former
$|h_3|\approx 0.063$, in the latter $|h_3|\approx 0.074$.  We also note
the average ratios $\langle h_3\rangle/\langle V_0/\sigma_0\rangle\approx
-0.12$ for one BH and $\langle h_3\rangle/\langle
V_0/\sigma_0\rangle\approx -0.15$ for two.  
These ratios are in excellent agreement with the empirical $h_3$ -- $\sigma/V$
relation of \citet{bsg94} in a sample of 44 elliptical galaxies.  
All galaxies in the Bender et al. sample show opposite signs of
$h_3$ and $V_0$ and fall near the relation 
$\langle h_3\rangle\approx -0.12 \langle V_0/\sigma_0\rangle$.  
The sudden sign change in $h_3$ very near the center is a consequence
of averaging over apertures;
similar features can be seen in the models of \citet{deh95} and
\citet{qia95}.

Interestingly, our simulations
present a direct counterexample to the results of \citet{bna01} whose
simulated mergers always yielded $h_3/(V_0/\sigma_0)>0$ in
apparent disagreement with the observations; 
these authors argue that the
presence of disk-like subcomponents may be necessary to reproduce the
correct sign of $h_3/(V_0/\sigma_0)<0$.  
The bulge initial conditions of
\citet{bna01} had $\rho\sim r^{-1}$ central cusps and were thus
significantly less concentrated than ours.  
New simulations of mergers
with a range of density profiles are needed to clarify the dependence of
$h_3$ on initial conditions.
\begin{figure*}[t]
\epsscale{2.0}
\plotone{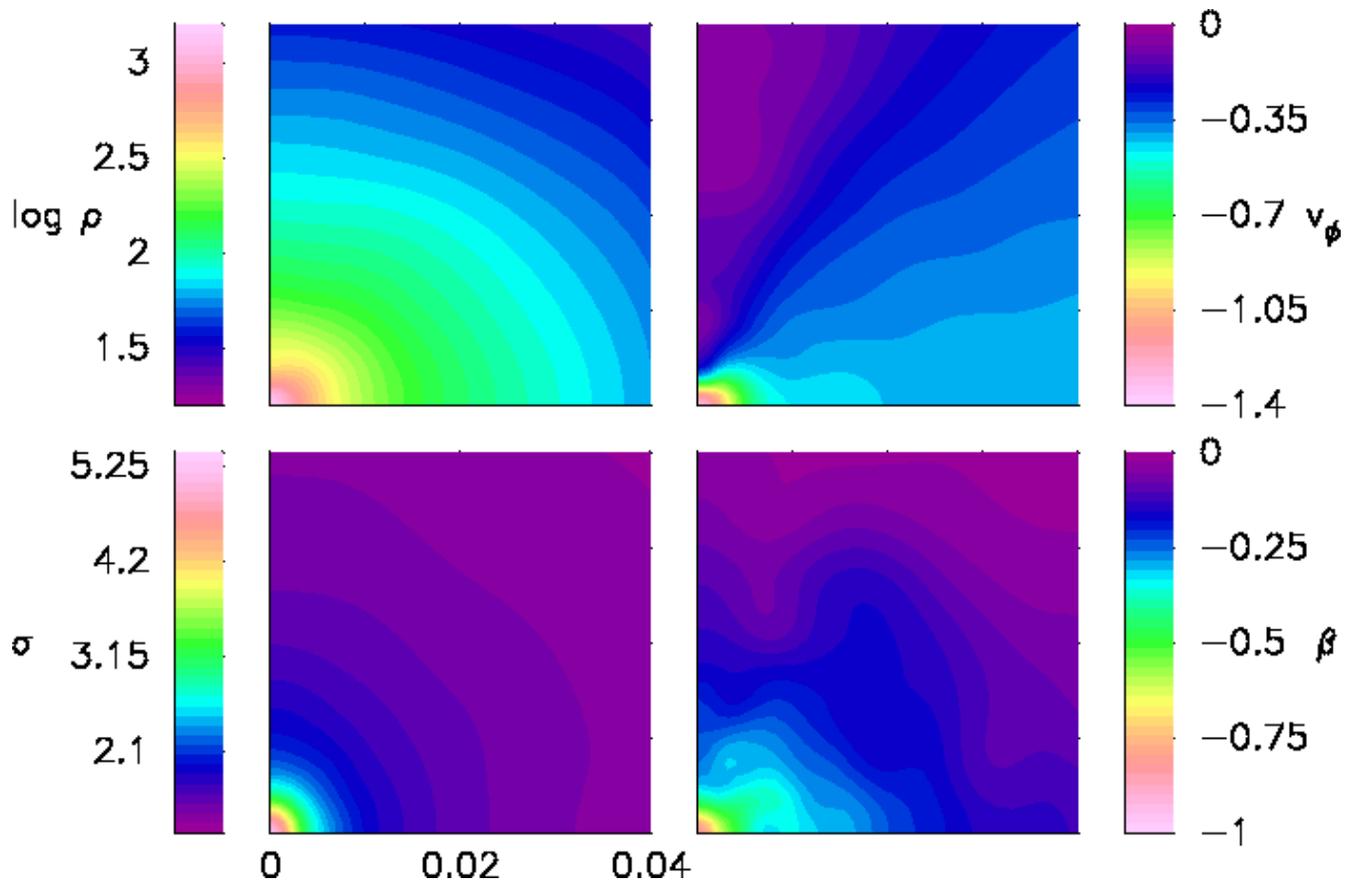}
\epsscale{1.0}
\caption{
Structure of the merger remnant in the meridional plane.  
The BH binary is at the lower left corner of each panel.  
Plots are averages over multiple snapshots as described for 
Figure \ref{fig_isophotes}.  
Horizontal axis is distance $\varpi$ from the $Z$ axis;
vertical axis is distance $Z$ from the equatorial plane.  
$\rho$ is the density of stars, 
$v_\phi$ is the average velocity through the meridional plane, 
$\sigma=\sqrt{\sigma_r^2/3+2\sigma_t^2/3}$ is the average 1D velocity 
dispersion, 
and $\beta\equiv 1-\sigma_t^2/\sigma_r^2$ is the anisotropy parameter, where $\sigma_r$ and $\sigma_t$ are, respectively, the radial and the tangential velocity dispersions.
\label{fig_meridional}}
\end{figure*}

4. {\it Fourth Gauss-Hermite moments.}
$h_4$ is very small except at the very center.  
At radii greater than twice the BH radius of
influence $r_{gr}$, our data are consistent with $h_4=0$.  
For $r_{gr}\lesssim r\lesssim 2r_{gr}$, $h_4$ dips into negative values;
this is again a consequence of averaging over finite apertures
\citep{deh95,qia95}.
For $r\lesssim r_{gr}$, $h_4$ becomes positive again with peak values
of about $0.027$ (one BH) and $0.083$ (two BHs).
Although it is tempting to ascribe this difference to the differing
effects of single and dual BHs, the Poisson uncertainties in
our determinations of $h_4$ are large enough that we are reluctant to 
infer a significant difference between the two models.

The new generation of 
integral field spectrographs like {\sf SAURON} \citep{pel01} 
can obtain two-dimensional maps of the stellar kinematics from 
absorption-line spectra at resolutions of less than an arcsecond,
corresponding to $\lesssim 1$ pc in nearby galaxies.
In Figure \ref{fig_gh}, we show 2D maps of the parameters 
$V_0$, $\sigma_0$, $h_3$ and $h_4$ similar to the maps obtainable with
{\sf SAURON}.
The line of sight is parallel to the plane of the merger,
which projects into the horizontal, $X$-axis of the image.  
To generate these images, we combined 100 snapshots in the time interval
$7.5<t<8.5$.  At every point on an $41\times41$ grid we started with an
aperture of diameter $D=0.01$ that was then shrunk to smaller $D$
under the condition that the number of stars inside the aperture 
always remain sufficient for the recovery of the LOSDVs 
($\textrm{a\ few}\times 10^3$).

The rotation pattern seen in projection is symmetric and co-aligned with the
merger's initial orbital axis (vertical axis in the image).  
The rotation reaches a maximum of about $|V_0|\approx 0.4$ 
at $r\approx 0.04$ on the major axis before dropping slightly at larger radii.
The velocity dispersion $\sigma_0$ exhibits no deviation from a circular
pattern in spite of the noticeably elliptical isophotes of the
density (Figure \ref{fig_isophotes}).
The third Gauss-Hermite moment $h_3$ is largest along the major axis
($h_3\sim\pm 0.1$ at $R=0.05$ is typical) and gradually decreases to zero
toward the minor axis.  
Its map resembles the rotation pattern of $V_0$, except
that the sign of $h_3$ is opposite from $V_0$ as noted above.
The fourth Guass-Hermite moment $h_4$ is consistent with zero 
everywhere except for the center; however there is a hint of
positive $h_4$ along the major axis and negative 
$h_4$ along the minor axis.

In the run {\sf B2} where the BH binary was replaced by single BH at 
$t=t_h$ (not shown here), the rotation field peaks at 
$V_0\approx\pm0.55$ much closer to the BH at
$|X|<0.01$ and remains high $|V_0|\gtrsim 0.5$ in a disk-like
circumnuclear region out to $|X|\sim 0.025$.  

Finally, in Figure \ref{fig_meridional} we present 2D maps of $\rho$, $v_\phi$, $\sigma$ and $\beta$ 
in the meridional plane, which is perpendicular to the merger plane and
contains the axis of rotation of the merger remnant.  
Here $\sigma\equiv\sqrt{\sigma_r^2/3+2\sigma_t^2/3}$ is the average
1D velocity dispersion and $\beta\equiv 1-\sigma_t^2/\sigma_r^2$ is the
anisotropy parameter.
In each panel, the
pixels are mapped in coordinates $(\varpi,Z)$, where $\varpi$ is the
distance from the axis of rotation and $Z$ is the distance above the 
equatorial plane (i.e. the plane of the merger).
We found symmetry
in all of the above variables under reflection $Z\rightarrow -Z$, therefore
only positive values of $Z$ are plotted.  As in Figure \ref{fig_gh},
the images were generated from a superposition of 100 snapshots in the
interval $7.5<t<8.5$, but this time we calculated non-parametric kernel
estimates for all quantities as described in Appendix \ref{ap_kernel}.  

\section{Discussion}
\label{sec_discuss}

\subsection{Scaling}

The scaling of our models to physical units will depend
on which time step we choose to compare with real galaxies
and on which galaxy we choose for comparison.
Fortunately there are two quantities that remain
nearly or exactly constant with respect to time in our simulations
and which are convenient for scaling:
the total mass $\mbin$ of the BH binary,
and the velocity dispersion $\sigma_*$ of stars in the nucleus,
outside the region where the stellar motions are strongly
affected by the BHs.
The first quantity is precisely constant,
while the second varies only slightly with time 
(e.g. Figure \ref{fig_binary}e) and position (Figure \ref{fig_moments}).

Furthermore there is a tight relation between BH mass and
stellar velocity dispersion in real galaxies, 
the $\ms$ relation, which allows us to reduce the scaling to 
a single number.
For our purposes, the most useful form of the $\ms$ relation is
\beq
M_{\bullet} = 1.30(\pm 0.36) \times 10^8 \msun \left({\sigma_c\over 200
\ \mathrm{km\ s}^{-1}}\right)^{4.72 (\pm 0.36)} 
\label{msig}
\eeq
\citep{mef01a}.
Here $\sigma_c$ is the projected velocity dispersion
measured in an aperture of radius $r_e/8$ centered on the BH,
with $r_e$ the half-light radius of the stars.
At ground-based resolutions, $\sigma_c$ is essentially unaffected
by the presence of the BH
and measures the velocity dispersion defined by the stellar
spheroid.
If we equate $\sigma_c$ with $\sigma_*$ in our definition for
$a_h$, the semi-major axis of a hard binary (equation \ref{def_hard}),
we find
\beq
a_h \approx 1.51\ {\rm pc} \left({\mh\over 10^8\msun}\right)^{0.576} .
\eeq

In our simulations, $\sigma_*(r_e/8)$ is close to $0.8$ in model
units at all times after formation of the hard BH binary.
The mass of the BH binary is $0.02$ in model units; 
we identify this with $\mh$,
the mass of the BH in the observed galaxy (or the combined mass of the
two BHs in the case of a binary).
If we define scaling factors 
\{${\cal M}$, ${\cal V}$, ${\cal L}$, ${\cal T}$\}
for our models, such that the mass in physical units is ${\cal M}$ times
the mass in model units and similarly for velocity, length and time, 
then (\ref{msig}) implies

\beq
0.02{\cal M} = 1.30\times 10^8\msun\left({0.8{\cal V}\over 200\ 
\mathrm{km\ s}^{-1}}\right)^{4.72}
\eeq
or
\beq
{{\cal M}\over 10^9\msun} = 2.27 \left({{\cal V}\over 200\ \mathrm{km\ s}^{-1}}\right)^{4.72}
\eeq
and
\beq
{\cal M} = 50\mh.
\eeq
The length and time scaling factors are
\begin{eqnarray}
{\cal L} &=& 390\ \mathrm{pc} \left({\mh\over 10^8\msun}\right)^{0.58} ,
\nonumber\\ 
{\cal T} &=& 1.62\times 10^6 \mathrm{yr} \left({\mh\over 10^8\msun}\right)^{0.37}.
\end{eqnarray}
We note that these scaling factors are independent of quantities like
the break radius $r_b$.
This is appropriate, since the empirical $\ms$ relation is also (apparently)
$r_b$- independent, and in our simulations, $\sigma_*$ (and hence ${\cal V}$)
are hardly affected by the creation of a core.

We consider two representative examples for the scaling.
Suppose we identify our simulations at early times 
(before cusp destruction) with a galaxy
like M32, a dwarf elliptical with a steep cusp.
The BH mass in M32 is $\mh\approx 3\times 10^6\msun$ 
\citep{jos00},
giving 
\begin{eqnarray}
{\cal M} &=& 1.5\times 10^8\msun,\ \ \ \ 
{\cal V} = 110\ \mathrm{km\ s}^{-1} , \nonumber\\
{\cal L} &=& 51\ \mathrm{pc}, \ \ \ \ 
{\cal T} = 4.4 \times 10^5 \mathrm{yr} .
\end{eqnarray}
The elapsed time in our simulations from $t=t_h$, the
time of formation of the hard binary, until the final time step
at $t\approx 20$ then corresponds to $\Delta t\approx 4.0\times 10^6$ yr.
The final value of $a$, the semimajor axis of the binary, is
$\sim 5\times 10^{-3}$ pc or $\sim 0.04 a_h$, 
and the final break radius is $\sim 3$ pc.

Or we could identify our simulations at late times
(after cusp destruction) with a galaxy like M87,
a bright elliptical with a weak cusp.
The BH mass in M87 is $\mh\approx 3\times 10^9\msun$ 
\citep{mac97},
which gives
\begin{eqnarray}
{\cal M} &=& 1.50\times 10^{11}\msun,\ \ \ \ 
{\cal V} = 490\ \mathrm{km\ s}^{-1} , \nonumber\\
{\cal L} &=& 2.8\ \mathrm{kpc}, \ \ \ \ 
{\cal T} = 5.7 \times 10^6 \mathrm{yr} .
\end{eqnarray}
The corresponding elapsed time is $\sim5.0\times 10^7$ yr,
the final value of $a$ is $\sim0.28$ pc (also $\sim 0.04 a_h$), and the
final break radius is $\sim170$ pc.  (This value is $\sim 3$ times
smaller than the break radius reported by \citet{lau92},
which probably means that M87 has undergone more than one major
merger since the era of formation of the supermassive BHs; 
see \S\ref{sec_deficit}.)

Henceforth we will refer to these as the ``M32'' and ``M87'' scalings
respectively.

\subsection{Cusps and Cores}

Our simulations demonstrate that the merger of two galaxies with 
steep, power-law density cusps ($\rho\sim r^{-2}$) can produce
a galaxy with a shallow power-law cusp ($\rho\sim r^{-1}$) inside 
of a break radius $r_b$; the necessary ingredient for the 
transformation is energy input from a pair of massive BHs.
Omitting the BHs \citep{bar99}, or artificially coalescing them
immediately after the merger (\S\ref{sec_merge}), preserves
the steep cusp.
As discussed above (\S\ref{sec_merge}),
most of the evolution of the central density profile in our simulations
takes place during a brief period when the two BHs first form a
hard binary.
Subsequent ejection of stars by the BH binary produces a gradual 
flattening of the inner slope (Figure \ref{fig_profs}).
The space density profile $\rho(r)$ is always well described as a power law,
$\rho\sim r^{-\gamma}$, at small radii.
However the projected density $\Sigma(R)$ looks qualitatively
different: a ``core'' appears, characterized by a log-log slope that
falls to zero at the center (Figure \ref{fig_profs}).
This is natural, since an $r^{-1}$ cusp in space density projects
to a logarithmic core in surface brightness (e.g. \citet{deh93}); 
only power laws steeper than $r^{-1}$ remain power laws
on projection.

Figure \ref{fig_profs} invites comparison with the luminosity
profiles of real galaxies.
These are well known to fall into two classes, the ``power laws''
and the ``cores,'' the latter having well-defined break radii.
It was initially argued \citep{lau95} that core profiles
were qualitatively distinct from power-law profiles, 
but \citet{mef96} pointed out the effect of projection
on a weak power-law cusp and used nonparametric deprojection techniques
to verify a power-law dependence of $\rho$ on $r$ even in the core galaxies.
More recent work \citep{geb96,res01} has verified 
this result in larger samples.
Our simulations demonstrate that nuclear density profiles with
a range of power-law slopes can be generated in a natural way 
starting from galaxies with steep cusps, and that the resultant nuclei
look like classical cores in projection when the index of the power 
law is small.

Surface brightness data for elliptical galaxies and bulges are commonly
fit to a parametric model, the ``Nuker'' law:
\beq
\Sigma(R)=\Sigma_0\xi^{-\Gamma}\left(1+\xi^{\alpha}\right)^{(\Gamma-\beta)/\alpha},\ \ \ \ \xi={R\over R_0}
\label{eq_nuk}
\eeq
\citep{lau95,byu96}.
This functional form (and the one due to \citet{fer94} on which it was based) 
has a built-in power-law dependence of $\Sigma$ on $R$ at small radii.
Our simulations preserve a power law in the {\it space} density
at small radii but not the {\it surface} density.
We suggest fitting power-law models like (\ref{eq_nuk}) to the
{\it de}projected density profiles of galaxies, rather than 
to their surface brightness profiles -- or equivalently, fitting
the projection of an expression like equation (\ref{eq_nuk})
to the surface brightness data.
If ``core'' galaxies really are characterized by weak inner power laws in
$\rho$, as in our models,
the fit to the data should thereby be improved.

\subsection{The $\mh-r_b$ and $\mh-M_{ej}$ Relations}
\label{sec_deficit}

Part of the motivation for putting core and power-law
galaxies into distinct categories was the observation that
core galaxies are systematically more luminous than power-law
galaxies.
Core galaxies have $-24\lesssim M_V\lesssim -20$ while power-law
galaxies are mostly fainter than $M_V= -20$ 
\citep{fer94,geb96,fab97},
although with considerable overlap at intermediate luminosities, 
$-22\lesssim M_V\lesssim -20.5$.
Recent studies (e.g. \citet{cas98,res01,rav01})
have confirmed this systematic difference
while weakening the case for a dichotomy; the variation of cusp slope
with galaxy luminosity is essentially continuous in the larger
samples now available.

What does our model predict?
Bright galaxies should have experienced more mergers than
faint galaxies and suffered more from the scouring action of
binary BHs.
This is the reasoning that led \citet{ebi91} to suggest that
the ``cores'' of giant ellipticals (which they took to be regions
of constant density, not weak power-law cusps) are generated by binary BHs.
\citet{fab97} showed that the ``core masses'' of bright ellipticals
scale with galaxy luminosity in roughly the same way  
as in the simulations of \citet{quh97}.
A point not made by these authors is that a pair of
BHs will eject of order its combined mass during {\it each} merger.
The total mass ejected depends both on the final BH mass, 
and on the number of stages in the merger hierarchy that have
occurred since the BHs first formed.
If the masses of the BHs at some stage in the merger hierarchy is $\mh/n$,
with $\mh$ the final BH mass,
the mass ejected by all the progenitors at this stage is of order 
$n\times\mh/n\approx \mh$,
and the total mass ejected in the complete set of mergers 
scales both with $\mh$ and with the number of mergers.
Since the latter is bigger for bigger galaxies, we expect 
to see a steeper-than-linear relation between
core mass and BH mass in elliptical galaxies.

To sharpen this argument and test it against real galaxies,
we need a working definition of ``core mass,'' or more precisely,
for the mass deficit -- the mass ejected by the BHs.
We define this as the mass needed to bring an observed density profile 
to a $\rho\sim r^{-2}$ dependence near the center.
Here we are assuming, as above, that $r^{-2}$ cusps were universally
present before the binary BHs began to do their damage and that they
would have been preserved in the absence of BHs.
This assumption (similar to the one made in \S\ref{sec_intro} when justifying
our choice of initial conditions) is reasonable since: 
(1) the growth of single BHs in pre-existing cores produces $\rho\sim r^{-2}$
cusps;
(2) steep cusps are preserved during mergers in the absence of
energy input from supermassive BHs (\S\ref{sec_merge});
(3) without BHs, pre-existing cores evolve into something like steep cusps 
through successive mergers \citep{mae96};
(4) faint elliptical galaxies universally have steep cusps.

Consider then a pair of galaxies with $\rho\sim r^{-2}$ central density cusps
which merge to form a galaxy with a shallower cusp,
$\rho\sim r^{-\gamma}$, $\gamma<2$, inside of a break radius $r_b$.
Assume that the density profile of the merger remnant was homologous
with that of the merging galaxies before the BHs began to
heat the stars.
The mass initially within $r_b$ was
\beq
4\pi\int_0^{r_b} dr\ r^2 {\sigma_*^2\over 2\pi Gr^2} = {2\sigma_*^2r_b\over G}
\eeq
and after mass ejection,
\beq
4\pi\int_0^{r_b} dr\ r^2 {\sigma_*^2\over 2\pi Gr_b^{2-\gamma}r^{\gamma}} = {2\over 3-\gamma} {\sigma_*^2r_b\over G}.
\eeq
We ignore changes in $\sigma_*$ (cf. Figure \ref{fig_binary}).
The ejected mass is therefore
\beq
M_{ej} \approx {2(2-\gamma)\over 3-\gamma} {\sigma_*^2r_b\over G}.
\label{eq_defmej}
\eeq

The break radius $r_b$ that appears in equation (\ref{eq_defmej}) 
refers to the space density $\rho(r)$,
while published break radii $R_b$ are derived from 
surface brightness profiles $\Sigma(R)$.
However the definitions of both $r_b$ and $R_b$ are to an extent 
arbitrary (e.g. \citet{fab97}) especially since we ignore in
our definition of $M_{ej}$ the precise form of the density
profile near the break radius; furthermore we find that $r_b\approx R_b$ 
within the uncertainties in our $N$-body models (Table \ref{tab_summary}).
We therefore feel justified in replacing $r_b$ by $R_b$ in
equation (\ref{eq_defmej}).
We find that $M_{ej}$ defined in this way is
larger by a factor $\sim 2$ than the ejected
masses that we measured in our simulations by counting stars 
that completely escape the galaxy (Figure \ref{fig_binary}f).  
There is a simple explanation for this discrepancy: 
after the ejection of one BH-binary-mass in stars,
the central gravitational pull on the remaining stars will decrease by 
roughly a half; 
these stars will then shift to wider orbits thereby increasing the 
apparent $M_{ej}$.
But this factor of $\sim 2$ is comparable to other uncertainties in
the definition of $M_{ej}$ and we will neglect it in what follows.

\begin{figure}[b]
\plotone{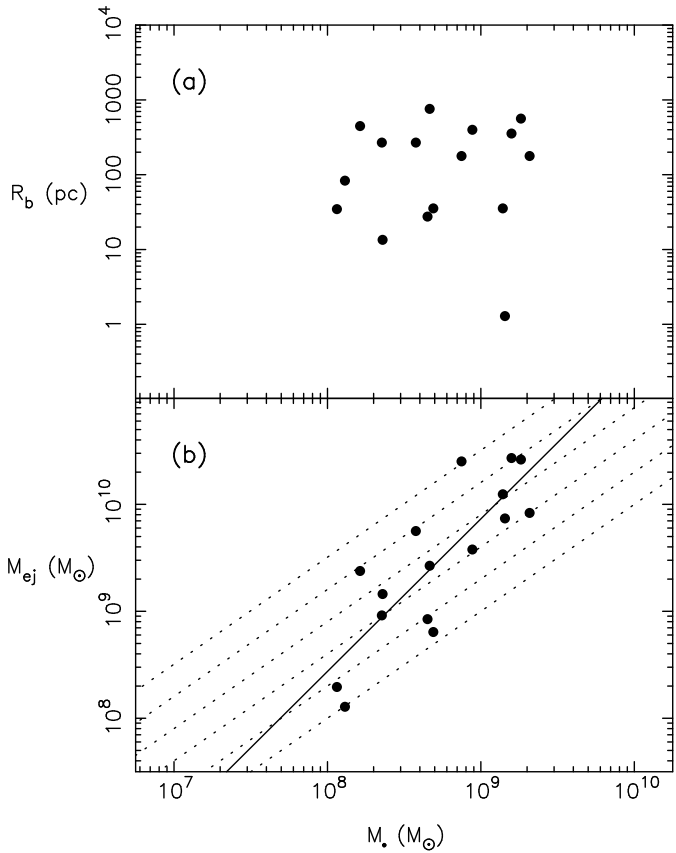}
\caption{
Correlation of BH mass with break radius (a) and ejected core mass (b) in
galaxies classified as ``core'' galaxies by Faber et al. (1997).
BH masses were computed using the $\ms$ relation (equation \ref{msig})
and measured values of the central velocity dispersion.
$M_{ej}$ is defined as in equation (\ref{eq_defmej}); the inner slope
$\gamma\equiv-d\log\nu/d\log r$ was taken from Gebhardt et al. (1996). 
Solid line in the lower panel is a least-squares fit.
Dotted lines show $M_{ej} = (1,2,4,8,16,32)\mh$.
\label{fig_rbsig}}
\end{figure}

\citet{fab97} list 16 ``core'' galaxies for which there are 
published measurements of $\gamma$, $R_b$ and $\sigma_*$.
Essentially none of these galaxies has an accurately-determined BH mass,
but we can use the $\ms$ relation (\ref{msig}) combined with measured
values of $\sigma_*$ to estimate $\mh$.
Figure \ref{fig_rbsig} shows the result.
The correlation of $M_{ej}$ with $\mh$ is reasonably tight,
and the slope of the relation is significantly greater than 
one, as predicted.
By contrast, $R_b$ and $\mh$ are essentially uncorrelated.
The brightest core galaxies ($M_V\approx -24$)
have $\langle M_{ej}\rangle\approx 10\mh$ while the faintest 
core galaxies ($M_V\approx -20$) have $M_{ej}\approx \mh$.
Ejection of $\sim 10\mh$ in stars in the biggest galaxies 
seems reasonable since these galaxies are believed to have
experienced several mergers since the quasar
epoch \citep{kcb01}.

While this success is encouraging, we point out that 
there are other factors that might contribute to the trend of
increasing $M_{ej}/\mh$ with galaxy mass.
The ``power-law'' galaxies have $\gamma\approx 2$ and
hence $M_{ej}\approx 0$ according to equation (\ref{eq_defmej}).
These galaxies would fall far below and to
the left of the ``core'' galaxies in Figure \ref{fig_rbsig}
and so it is reasonable to expect a steeper-than-unit slope for
the brighter galaxies that are plotted there.
We return in the next section to the question of how the power-law galaxies
managed to maintain steep cusps in the presence of mergers.

\subsection{Persistence of Steep Cusps in Faint Galaxies}

The story just outlined can not be complete, since
steep power-law cusps persist in elliptical galaxies as
bright as $M_V\approx -22$ \citep{geb96},
and are universally present in galaxies fainter than $M_V\approx -20$.
How have the cusps in the power-law galaxies managed to avoid
destruction by merging BHs?
We discuss several possibilities.

1. {\it Power-law galaxies do not contain supermassive BHs.}
While the presence of supermassive compact objects has only
been reliably established in a handful of galaxies 
(see discussion in \citet{fem00}), several of the best
cases are in stellar systems with steep cusps (e.g. the bulge of the
Milky Way; M32).

2. {\it Power-law galaxies were not formed via mergers.}
This would contradict standard models for hierarchical structure
formation \citep{lac93,hak00,mhn01}, although it is possible
that most power-law galaxies have not experienced major mergers since
the era when BHs gained most of their mass; we return to this idea below.

3. {\it Cusps in power-law galaxies are regenerated following mergers.}
\citet{fab97} suggested that steep cusps might be produced by star 
formation from fresh gas supplied during mergers.
While mergers certainly lead to enhanced star formation,
we consider this explanation unlikely since nuclei formed from
infalling gas do not resemble featureless power laws 
(e.g. \citet{mih94}).
The scale-free nature of the cusps suggests to us a  
gravitational origin.

4. {\it A mechanism exists for extracting energy from BH binaries 
before they can heat the surrounding stars.} 
This suggestion is motivated by the observation that the central
density profiles in our simulations remain homologous with the 
initial profile, $\rho\sim r^{-2}$, for a short time after the
merger (Figure \ref{fig_homol}).
If some process were effective at extracting the binary's energy at 
this stage, at a rate higher than the hardening rate due to stellar
ejection, cusp disruption could be avoided, producing a coalesced
BH binary in a steep cusp.
We tested this idea in a limited way above (\S\ref{sec_merge})
by artificially combining the two BHs immediately after the merger;
the profile retained its steep power-law character thereafter.

We propose that explanation (4) is the correct one and that the 
mechanism which extracts energy from the BH binary is gas dynamical 
in origin.
The effects of gas on the evolution of BH binaries 
have been discussed by a number of authors.
Gas may accrete onto the larger of the BHs causing orbital 
contraction at a rate $\sim\mbin/\dot M$ \citep{bbr80,val96}.
Interaction of the binary with an accretion disk will
transfer angular momentum from the BHs to the gas causing the 
binary orbit to decay \citep{lip79a,lip79b,syc95,arl96};
the rate is again of order the gas accretion rate
\citep{ipp99}.
Discrete gas clouds, like those near the center of the Milky Way
\citep{coh99}, might also affect the evolution of the
binary, particularly when its separation is large.
Gas clouds could scatter stars into the BH binary's
loss cone \citep{kim01},
enhance the Brownian motion of the binary,
or even perturb the individual BH trajectories 
before they form a bound pair \citep{bek00}.

A difficulty with explanations that demand gas accretion
is the high accretion rate implied:
a mass of order $\mbin$ must find its way to the nucleus
on the time scale of the merger,
or roughly one crossing time,
so that the binary avoids disrupting the stellar cusp.
Much of the energy released from the BH binary
would go into heating the gas.
It follows that the events we are envisioning are energetically
similar to quasars, a conclusion reached also by 
\citet{gra90}, \citet{gor00} and others.
If this picture is correct, 
power-law galaxies acquired most of their mass before and
during the quasar epoch, from progenitors that were gas-rich,
while core galaxies are ellipticals whose most recent major merger 
occurred after the era of formation of the BHs.

There is circumstantial evidence in support of this picture.
The division between core and power-law galaxies occurs at
$-22\lesssim M_V\lesssim-20.5$ \citep{fab97}.
In semi-analytic models for galaxy formation,
the predicted ratio
of gas mass to stellar mass during the last major merger
is a steep function of galaxy luminosity, dropping from
$\sim 3$ for $M_V=-18$ to $\sim 0.3$ for $M_V=-21$ 
\citep{kah00}.
Thus the gas content of the progenitors of the
power-law galaxies was likely to have been high 
at all previous stages in the merger hierarchy, 
while for core galaxies the most recent mergers were 
probably gas-poor.
Furthermore the redshift of the last major merger 
(defined as a merger with mass ratio less extreme than $1:3$)
is a strong function of a galaxy's current mass.
Most galaxies with masses less than $\sim 10^{10}\msun$
($M_V\approx -18$) have never experienced a major
merger; only galaxies with $M\gtrsim 10^{11}\msun$ ($M_V\approx -21$)
have typically undergone a major merger since a redshift of $1$
\citep{kcb01}.

Sharpening these arguments will require $N$-body simulations
of BH binary evolution including gas.

\subsection{Coalescence Time Scales and Persistence of Binary BHs}

We identify two characteristic time scales associated with the decay 
of the BH binary in our simulations. 
The first, which is essentially $N$-independent, 
is the brief period following the merger when the two BHs fall to the
center and form a hard binary.
As discussed above (\S\ref{sec_merge}), the BHs initially come 
together in a time that is approximately as long as the merger itself; 
most of the energy transfer from the BHs to the stars in the 
nucleus takes place in just $\Delta t\approx 0.1$ in model units (Figure 
\ref{fig_lagrange}), or $\sim 10^5-10^6$ yr.
The separation between the BHs at the end of this period, $t\approx t_h$,
is $\sim 10^{-3}$ in model units (Figure \ref{fig_gdtnb6}) corresponding
to roughly $0.05$ pc (M32) or $3$ pc (M87).

The second time scale is associated with the gradual decay of the BH 
binary.
We especially wish to know how long it will take the binary to 
decay to the point that emission of 
gravitational radiation becomes the dominant energy sink.
This occurs when $|\dot a/a|^{-1}$ due to mass ejection first 
equals $t_{gr}$ as 
defined above (equation \ref{eq_tgrav}).
This second time scale is potentially $N$-dependent,
since $|\dot a/a|$ depends to some extent on collisional 
processes which may be much larger in our simulations than
in real galaxies (\S\ref{sec_hard}).

For the moment, assume that the decay rate of the BH binary in 
our simulations is characteristic of real galaxies with much 
larger $N$.
The inverse semimajor axis increases roughly linearly with time
in our runs (Figure \ref{fig_binary}b), which suggests that we
define a decay rate
\beq
S\equiv {d\over dt}\left({1\over a}\right).
\eeq
In model units, $S\approx 7.0\times 10^2$; in physical units,
\beq
S \approx 1.0\times 10^{-6}\left({\mh\over 10^8\mh}\right)^{-0.95} {\rm yr}^{-1} {\rm pc}^{-1}.
\label{sphys}
\eeq
We explore the consequences of assuming that $S$ remains constant
at this value until gravity-wave coalescence takes place,
then discuss the reasonableness of this assumption.

Energy loss due to gravitational radiation dominates that from
stellar interactions when
\beq
t_{gr} = {1\over aS}
\eeq
with $t_{gr}$ given by equation (\ref{eq_tgrav}),
or
\beq
a^5={64\over 5S}{G^3\mbin^3\over c^5}
\eeq
and $F(e)$ (equation \ref{fofe}) is henceforth set to one.
(The modest rates of growth of $e$ in our simulations,
\S\ref{sec_hard}, imply $F\approx 1$ at all but the latest
stages of the merger, and $a\propto F^{-0.2}$.)
Combined with equation (\ref{sphys}),
this condition becomes
\beq
a<a_{crit}\approx 0.012\ {\rm pc}\left({\mh\over 10^8\msun}\right)^{0.8}.
\eeq
Scaling to real galaxies we find
$a_{crit}\approx 8\times 10^{-4}$ pc (M32)
and $\sim 0.2$ pc (M87).
These values are smaller, by respective factors of $\sim 0.12$ and
$\sim 0.6$, than the final value of $a$ in our simulations.
The gravitational radiation time scale when $a=a_{crit}$, 
which is also approximately equal to the time elapsed in 
reaching $a_{crit}$ from $a_h$ in this simple model, is
\beq
t_{gr}(a_{crit}) \approx 5.0\times 10^7{\rm yr}\ \left({\mh\over 10^8\msun}\right)^{0.16}
\eeq
which is $3.0\times 10^7$ yr (M32) and $9.0\times 10^7$ yr (M87).
These are factors of $\sim 8.0$ (M32) and $\sim 1.8$ (M87) longer
than the elapsed time from $t=t_h$ until the final time step
in our simulations.
Thus a straightforward extrapolation of our $N$-body results implies
that BH binaries would achieve gravitational radiation coalescence in
a relatively short time, of order $10^8$ yr, following a merger.

Is this a reasonable conclusion?
Although the decay rate of the BH binaries in our simulations showed
no appreciable $N$-dependence over the range of $N$ that we tested
(Figure \ref{fig_binary}b),
other aspects of the evolution were observed to depend strongly on $N$,
including the amplitude of the Brownian wandering 
(Figure \ref{fig_wander}).
We argued above (\S\ref{sec_hard}) that efficient decay of the
binary in our simulations depended on this wandering,
since it allowed the binary to interact with a larger pool of stars
than if it remained precisely fixed at the center.
We predicted that the supply of stars would have been exhausted
by the end of our simulations if the wandering had been reduced by
increasing $N$ to $\sim 5\times 10^5$, still much smaller than in
real galaxies.

In a real galactic nucleus,
the random velocity of a supermassive BH is expected to be only
\begin{eqnarray}
v_w &\equiv& \sqrt{\langle v^2\rangle} \nonumber\\
&\approx& 0.033\: {\rm km\ } {\rm s}^{-1}\left({m_*\over\msun}\right)^{1/2} \left({\mh\over 10^8\msun}\right)^{-0.29};
\end{eqnarray}
the $\ms$ relation has been used to express $\sigma_*$ in terms of
$\mh$.
(This velocity would be increased by a modest factor, of order unity,
for a binary BH with $\mbin=\mh$
(\citet{mer01}; Figure \ref{fig_equipart}).)
The corresponding wandering radius is
\beq
r_w \approx 0.010 \: {\rm pc}\left({\mh\over 10^8\msun}\right)^{-2.35}\left({r_c\over 100\ {\rm pc}}\right)
\label{wander}
\eeq
with $r_c$ the core radius inside of which the stellar density is 
taken to be constant.
These relations are based on a simple model \citep{mer01} in which
the stellar distribution is assumed to be unaffected by the motion of the
BH or BH binary.
If we accept these figures, the wandering radius of a BH binary
in a real galaxy is of order or less than the semimajor axis of the 
binary.
It follows that Brownian motion can generally be neglected when
considering the interaction of the binary with surrounding stars.

This conclusion may be too pessimistic.
A nearly stationary binary would soon ``scour clean'' a nearly-spherical 
region of radius $\sim 2a-3 a$ \citep{zie00}, 
after which the force acting on it
would be essentially zero and even a very small $v_w$ would translate
into a large displacement.
The amplitude of the Brownian motion might therefore slowly increase
with time as the binary eats its way through the nucleus.
Our simulations tell us nothing about the plausibility of this scenario
since our runs never reached the point of loss-cone depletion.
Larger $N$, longer runs, or initial conditions with lower central
densities would be required to test this idea.

But suppose that the Brownian motion remains always small,
with an amplitude less than the separation between the BHs.
We can conseratively estimate the final hardness achieved by the
binary by the following argument.
Assume that the center of mass of the binary remains fixed,
and that the shrinking binary ejects stars in order of their pericenter,
from smallest to largest.
In this way, the binary acts to reduce the central
stellar density in the most rapid possible way, causing
its decay to stall in the minimum time.
This process would create a hole at the galaxy's center which
grows as the binary shrinks;
at some point the binary lies entirely within the hole and
its decay ceases.
Using Figure \ref{fig_peri}, it is easy to show that the critical
separation is $\sim$ a few times $10^{-4}$ (assuming that
the decay stalls when the hole grows to a radius of a few times $a$).
Scaling to M32, this separation is $a\approx 0.02$ pc, 
and to M87, $a\approx 1$ pc.
The gravitational radiation time scale at these separations would
be $10^{13}$ yr (M32) and $10^{11}$ yr (M87), 
much longer than the age of the universe.
In reality, much of the ejected mass in the early stages of the
decay comes from stars with larger pericenters, allowing the
binary to shrink more than this, say by a factor of $\sim 2$.
(This is, coincidentally, roughly the final separation reached
in our simulations.)
However the basic conclusion is unchanged.
In galaxies with initially shallower cusps the decay would 
be expected to stall at still larger separations.

The inefficiency of stellar-dynamical processes at bringing
together supermassive BHs has been noted by a number of
authors (e.g. \citet{por94,val96,mer00,gor00}).
We believe that the problem has often been overstated due to the
use of over-simplified models for describing the interaction
of stars with the binary;
for instance, \citet{gor00} ignore the fact that most of the
hardening comes from stars on orbits with apocenters much greater
than the semimajor axis $a$.
But the basic argument is sound:
a {\it fixed} binary would have difficulty interacting with 
several times its own mass in stars even if located at the center of
a steep density cusp, hence it could probably not achieve 
gravitational radiation 
coalesence in a time shorter than the age of the universe.
Unless additional mechanisms exist for extracting energy,
its decay would be expected to stall.

We discussed above what these ``additional mechanisms'' might be and
argued that they would be most effective in galaxies whose progenitors
were gas-rich.
These galaxies, which in our opinion are {\it least}
likely to contain BH binaries, are also the systems in which detection of
BH binaries would be easiest, through the measurement of
periodically varying features in the emission line 
systems associated with one or both of the BHs 
\citep{bbr80,gas95}.
Such features have tentatively been detected in a few active 
galaxies; the best case is probably OJ 287,
a blazar with nearly-periodic outbursts dating back roughly $100$ years
\citep{pur00}.
However the interpretation in terms of a binary system 
\citep{lev96} is not airtight.
A larger number of interacting systems exhibit emission from
two resolved peaks, probably the nuclei of galaxies in the early
stages of merging (e.g. \citet{car90});
the smallest projected separation, in ARP 220, is $\sim 360$ pc
\citep{sco98}.
A recently-discovered double quasar contains two peaks with a 
projected separation of a few kiloparsecs \citep{jun01}.

The supermassive BHs in these interacting systems have not yet formed 
bound pairs.
True BH binaries -- at separations $a\lesssim a_h$
(equation \ref{def_hard}) -- would most likely be found in galaxies
with low central densities and little gas.
Recent formation via mergers, and a high ongoing accretion rate 
(assuming that the accreted galaxies also contain BHs),
would also be propitious.
These characteristics constitute almost a textbook definition of a cD galaxy,
particularly a multiple-nucleus cD in a rich galaxy cluster 
(e.g. \citet{scg82}).
Most cD galaxies are too distant for single, much less double, BHs to
be detected kinematically, although
a strong case can be made for dual BHs in 3C75, 
the central radio source in A400.
This galaxy exhibits a pair
of radio jets that appear to be emitted from point sources separated 
by $\sim 7$ kpc in projection \citep{owe85}.
But a more likely separation for a binary BH would be the much smaller
distance at which the decay is expected to stall,
$0.01\ {\rm pc}\lesssim a\lesssim 1$ pc.
Binaries with these separations might barely be detectable in
nearby galaxies by using VLBI techniques to resolve the compact
radio sources associated with the individual BHs \citep{sle94}.

The existence of uncoalesced BH binaries in galactic nuclei 
has a number of interesting consequences.
BH ejections would result whenever a third supermassive BH, or a second
BH binary, is brought into the nucleus following a merger 
\citep{val96}.
In a massive galaxy like a cD, a quasi-steady state might be set
up in which the ejection of BHs from the nucleus is matched by
the infall of new BHs from the ``multiple nuclei'' and from
previous ejections.
If BH ejections are common, most supermassive BHs might be located
far from the centers of galaxies, and the mean mass density of 
BHs in the universe could be much greater than the mass density
inferred from nuclear kinematical studies \citep{mef01b}.
 
\subsection{Centers of Dark-Matter Halos}

Cold dark matter (CDM) simulations of the growth of structure in the 
universe  \citep{moo98,jin00,bul01} predict dark-matter halos with 
steep central density cusps,
$\rho\sim r^{-\gamma}$, $1\lesssim\gamma\lesssim 2$, 
similar to the cusps in our initial models.
The dense (baryonic) regions in which supermassive BHs first formed were 
probably located near the centers of these halos
\citep{hnr98}.
Cosmological simulations currently lack the resolution to handle compact
massive objects like BHs, but a number of authors 
\citep{ips87,gos99,uzk01}
have investigated the response of pre-existing dark matter halos to the 
growth of supermassive BHs.
An initially steep dark-matter cusp becomes even steeper
within the radius of influence $r_{gr}$ of the BH,
$\rho_{DM}\sim r^{-A}$, $2.25\leq A\leq 2.5$ \citep{gos99}.
This result has been claimed to be inconsistent with experimental
upper bounds on annihilation radiation from the Galactic center
\citep{gos99,bss00,gon00},
implying either that dark matter cusps do not exist, 
or that current ideas about the composition of the dark matter 
are wrong.

If mergers of dark-matter halos occurred after the BHs were in place,
however, the effect of the BHs on the dark matter density would be roughly the 
opposite of what these studies assume: the BHs would tend to destroy the cusps
via ejection of dark matter particles.
Just the first step in the BH merger process -- formation of a hard binary via 
dynamical friction -- transfers enough energy to the background to convert 
an $r^{-2}$ cusp into a shallower, $\sim r^{-1}$ cusp within a radius 
that contains several times the BHs' mass (\S\ref{sec_merge};
Figure \ref{fig_profs}).
(This conclusion is independent of any uncertainties about Brownian
motion, loss-cone refilling, etc., which would be negligible
anyway in a dark-matter-dominated cusp.)
Most of the annihilation radiation from a putative dark-matter
cusp would come from a region smaller than this;
reducing the cusp slope from $\sim -2$ to $\sim -1$ lowers
the predicted flux by several orders of magnitude (\citet{gos99},
Fig. 2).
The fact that steep cusps persist in the {\it stellar} density
in many galaxies, including the Milky Way \citep{ale99},
suggests that dark matter cusps might also sometimes avoid destruction.
But the mechanisms discussed above for preserving
steep cusps in the stellar distribution -- e.g. star formation from
infalling gas -- are less applicable to dark matter.
Furthermore we expect the coupling between baryons and dark matter to be  
less than perfect, and stellar cusps would themselves inject energy into 
the dark matter as they spiralled to the center of the merging dark matter 
halos; a similar effect is seen in the merger simulations of \citet{mec01}.
Thus it seems possible that dark matter cusps could be destroyed
even in galaxies which manage to retain steep cusps in the stellar
distribution.

Could the mechanisms discussed in this paper be relevant to the
low apparent density of dark matter at the centers of dwarf 
and low-surface-brightness galaxies \citep{flp94,dem97,mcd98,deb01}?
The density increases only as 
$\rho_{DM}\sim r^{-0.2}$ at the centers of these galaxies \citep{deb01},
much flatter than predicted by CDM models.
It is unlikely that BHs alone are responsible for this deficit, 
however, for several reasons.
There is currently no evidence for supermassive BHs in these galaxies,
and scaling relations like the $\ms$ relation would suggest small values of
$\mh$.
The inferred core radii of the dark matter halos are very large,
of order $10^2-10^3$ pc, implying an ejected
mass that is much greater than any likely value of $\mh$.
Dwarf and low-surface-brightness galaxies are also unlikely to have 
had active merger histories,
at least since the era of formation of the BHs.
The only possibility we can see for destruction of dark-matter
cusps on the observed scales in these galaxies would be the existence 
of a significant population of condensed objects that predate the dark 
matter halos, such as primordial BHs \citep{car85}.

\section{Conclusions}
\label{sec_conclude}

1. Mergers of equal-mass stellar systems containing supermassive
black holes and steep central density cusps produce nuclei with shallow 
cusps, $\rho\sim r^{-1}$, inside of a break radius $r_b$,
similar to the luminosity profiles observed at the centers of
bright elliptical galaxies.
Most of the evolution in the central density occurs within a 
short time, $\sim 10^6-10^7$ yr, after the black holes form a binary;
the cusp continues to flatten thereafter as the binary
ejects stars via the gravitational slingshot.
The dependence of core properties on black hole mass in observed
galaxies is shown to be consistent with this formation model.

2. The merger-induced rotation in the nucleus is reduced
significantly by the binary as it preferentially
ejects stars whose angular momenta are aligned with its own.
The stellar velocity dispersion tensor in the nucleus
becomes mildly tangentially anisotropic as well, 
although this effect is too small to be easily observed
in real galaxies.

3. Hardening of the black-hole binary takes place efficiently in our 
simulations due to the large supply of stars provided by the dense cusps,
and also to the Brownian motion of the binary, which allows it to interact
with a larger number of stars than if it remained fixed.
There is no significant dependence of the binary hardening
rate on number of particles $N$ and the binary's loss cone
never approaches depletion,
in spite of the fact that the decay is followed over a
factor of $\sim 20$ in semimajor axis after it first forms a
hard binary, considerably farther than in earlier simulations.
The hardening rate that we measure, if it remained constant,
would imply gravitational-radiation coalescence in a 
relatively short time, of order $10^8$ years, following the merger.

4. However, we argue that the decay of a black-hole binary in a
real galactic nucleus would sometimes be expected to stall at separations 
of $0.01- 1$ parsec due to depletion of the stellar loss cone
around a nearly-stationary binary.
At these separations,
gravitational radiation would be ineffective at inducing coalescence
and the binary would persist indefinitely,
unless some other physical process were able to extract its binding energy.
We argue that uncoalesced black-hole binaries are most likely 
to be found in the nuclei of cD or other giant elliptical galaxies.

5. If we artificially combine the two black holes just after they form
a hard binary, the merger remnant preserves its steep,
$\rho\sim r^{-2}$ density cusp.
We propose this as a model for the retention of steep cusps
in the ``power-law'' galaxies,
and suggest that these galaxies 
experienced their last major mergers during the quasar epoch.

6. Our simulations can also be interpreted as describing mergers
of dark-matter cusps containing supermassive black holes.
We argue that the steep cusps predicted by cold-dark-matter
cosmologies would be destroyed by binary black holes
in galaxies where the stellar cusps are also destroyed.

\bigskip\bigskip

\acknowledgments

This work was supported by NSF grants AST 96-17088 and 00-71099 and by
NASA grants NAG5-6037 and NAG5-9046.
We thank Sverre Aarseth, Marc Hemsendorf and Rainer Spurzem for their
patient and expert guidance with the $N$-body codes {\sf NBODY6\ } and \NB,
and for making their programs available to us in advance of general release.
Without their generous help this project would not have been possible.
We discussed the potential observability of binary supermassive
black holes with E. Sadler and J. Wrobel.
The pre-hard-binary merger simulations presented here (\S2) were
carried out by Fidel Cruz using the tree code \GDT on the Rutgers
HPC-10000 supercomputer;
we are indebted to him and to Volker Springel for advice about 
using this code.
Dr. Cruz was supported by a fellowship from
the Consejo Nacional de Ciencia y Tecnologia de Mexico.
This work was partially supported by the National Computational 
Science Alliance under grant no. MCA00N010N and utilized the 
San Diego Supercomputer Center Cray T3E.
We are also grateful to the Center for Advanced Information Processing
at Rutgers University for their generous allocation of computer time
on the HPC-10000.

\onecolumn
\appendix

\section{The Coulomb Logarithm}
\label{ap_lnlambda}

Estimates of the orbital decay rate due to dynamical friction in
\S \ref{sec_merge} were dependent on the Coulomb logarithm
$\ln\Lambda$.
Here we derive estimates for $\ln\Lambda$ in the two cases of interest:
a single massive object near the center of a stellar system
with a steep density profile; and a sphere of finite size representing
a merging cusp.
In both cases we find $\ln\Lambda\approx 1$.

The deceleration of a massive test particle due to dynamical friction is often written in the form (e.g. \citet{cha43}; \citet{spi87})
\begin{equation}
\label{eq_dynf}
\langle\Delta v_\parallel\rangle = - \frac{4\pi G^2M\rho \ln\Lambda F(v)}{v^2}
\end{equation}
where $M$ and $v$ are the mass and the velocity of the test particle, 
$\rho$ is the density of light field particles, 
$F(v)$ is the fraction of field particles with velocities less than $v$, 
and $\ln\Lambda$ is the Coulomb logarithm that arises in the integration 
over a field of constant density.  
The force exerted on the test particle by an infinite homogeneous field 
diverges and a cutoff in the form of a maximum impact parameter $p_{max}$ 
is required.  
The dependence of $\langle\Delta v_\parallel\rangle$ on $p_{max}$ is 
logarithmic:
\beq
\langle\Delta v_\parallel\rangle\propto \ln\sqrt{1+\frac{p_{max}^2}{p_{min}^2}}
\equiv \ln\Lambda
\label{eq_df1}
\eeq
where $p_{min}$ is a minimum impact parameter cutoff that also needs to be specified in context.  
In treatments where the field particles are assumed to move along 
rectilinear orbits (e.g. \citet{cvn42}),
the integral leading to (\ref{eq_df1}) diverges at low impact parameters.
The divergence vanishes when the proper Keplerian trajectories are used;
field stars of a given relative velocity $V_0$ then transmit a net momentum
proportional to $\ln\sqrt{1+p^2_{max}/p_0^2}$ with $p_0=GM/V_0^2$,
and integration over $V_0$ gives a $p_{min}$ in equation (\ref{eq_df1})
of order $GM/\sigma^2$ with $\sigma$ the 1D velocity dispersion of the
field stars.
For instance, when the velocity $v$ of the test star is much less than
$\sigma$, $p_{min} \approx GM/\sqrt{2}\sigma^2$,
roughly the radius of gravitational influence $r_G$ of the massive object
\citep{mer01}.

A variety of prescriptions can be found in the literature regarding the 
optimal and most accurate choice for $p_{max}$, 
and thus for $\ln\Lambda$. 
Since all gravitationally-bound structures in the universe have finite extent, 
the Coulomb logarithm is in theory a real physical quantity subject to 
calculation if one is ready to abandon several simplifying assumptions that 
enter equation (\ref{eq_dynf}).  
In practice, the full-fledged phase-space integration is cumbersome at best
and numerical $N$-body treatments often resort to the fitting of equation 
(\ref{eq_dynf}) to the dynamical drag observed in simulations.

\citet{mao93} derived an expression for dynamical friction in an 
inhomogeneous isothermal Maxwellian background which reads
\begin{equation}
\label{eq_dfmaoz}
\langle\Delta v_\parallel\rangle=
-\frac{\sqrt{2}G^2 M}{v\sigma}
\int d^3 r \frac{\rho({\bf r})\alpha}{\left|{\bf R}-{\bf r}\right|^3}
\left[e^{\alpha^2-x^2}\erf(\alpha)-1\right]
\Theta\left(|{\bf R}-{\bf r}|-d\right)
\end{equation}
where $\alpha\equiv{\bf x}\cdot({\bf R}-{\bf r})/|{\bf R}-{\bf r}|$ and ${\bf x}\equiv{\bf v}/\sqrt{2}\sigma$, while ${\bf R}$ is the position of the test particle and $\Theta(y)=1$ when $y>0$ and is zero otherwise.  The $\Theta$-function serves to exclude a finite small volume of radius $d$ around the particle from integration, necessary since in Maoz's treatment the field-star trajectories
are assumed to be straight lines.

If the stellar system is spherical and centered on the test particle,  the radial and the angular integrals in equation (\ref{eq_dfmaoz}) can be separated
\begin{equation}
\langle\Delta v_\parallel\rangle=-\frac{\sqrt{2}G^2 M}{v\sigma}\left(\frac{2\pi}{x}\right)
\left\{e^{-x^2}\int_{-x}^x\alpha e^{\alpha^2}\erf(\alpha)d\alpha\right\}
\left\{\int_d^\infty \frac{\rho(r)}{r} dr\right\}.
\end{equation}
The first factor in braces can be identified with the velocity factor $F(v)$ appearing in equation (\ref{eq_dynf}) 
\begin{equation}
e^{-x^2}\int_{-x}^x\alpha e^{\alpha^2}\erf(\alpha)d\alpha
=\erf(x)-x\erfp(x)\equiv F(v)
\end{equation}
The second factor in braces encapsulates the dependence of dynamical friction on the radial distribution of field particles.  The formula becomes
\begin{equation}
\langle\Delta v_\parallel\rangle=
-\frac{4\pi G^2 M F(v)}{v^2}
\int_d^\infty \frac{\rho(r)}{r} dr
\end{equation}
which can be compared with equation (\ref{eq_dynf}) to arrive at a definition of the Coulomb logarithm in terms of an arbitrarily chosen fiducial density $\rho$:
\begin{equation}
\label{eq_lambdadef}
\rho \ln\Lambda \equiv \int_d^\infty \frac{\rho(r)}{r} dr .
\end{equation}
Clearly, Maoz's formula reduces to equation (\ref{eq_df1}) if the density is 
constant in an annulus with inner and outer radii $d$ and $p_{max}$ and 
vanishes outside.  

Real stellar systems do not have large-radius density cutoffs but the 
density typically decays as a power law $\rho\sim r^{-\lambda}$.  
The spatial integral in equation (\ref{eq_dfmaoz}) converges for {\it any}
$\lambda>0$, and one is free to take the limit $p_{max}\rightarrow\infty$.  
To illustrate this, we calculate the contribution due to the far field 
($r\geq d$) particles obeying the simplest power-law profile centered on 
the test particle:
\begin{equation}
\rho(r) = \rho_0 \left(\frac{r}{d}\right)^{-\lambda}.
\end{equation}
Assume for the moment that $\rho(r)=0$ when $r<d$, 
reflecting a ``hole'' in the density cusp due to, e.g., 
loss-cone depletion around a compact massive object.
In this case the Coulomb logarithm must be chosen as
\beq
\rho\ln\Lambda=\rho_0\int_{d}^\infty  \left(\frac{r}{d}\right)^{-\lambda} \frac{dr}{r}= \rho_0\frac{1}{\lambda} = {\rho(d)\over\lambda}
\label{eq_lnl}
\eeq
leading to dynamical friction that is proportional to the density at the 
hole's inner edge and inversely proportional to the logarithmic slope:
\beq
\label{eq_dfcusp}
\langle\Delta v_\parallel\rangle=
-\frac{4\pi G^2 M F(v)\rho(d)}{\lambda v^2}.
\end{equation}
This result helps expose the inadequacy of conclusions drawn in the context 
of homogeneous backgrounds where greatest contribution to the drag force 
comes from distant encounters ($r\gg d$).  
In our case, 
the fractional contribution from near-field particles at distances 
$d\leq r\leq 2d$ amounts to $1-2^{-\lambda}$ which is more than $50\%$ 
when $\lambda>1$.  
Also, while in view of the traditional choice $\ln\Lambda=p_{max}/p_{min}$ 
one is prone to expect $p_{min}\ll p_{max}$, 
we find that $p_{max}/p_{min}\sim e^{1/\lambda}\sim 1$ and any meaningful 
choice for the effective $p_{max}$ would reflect neither Chandrasekhar's 
``average distance between stars'' nor the ``size of the system.''

In the absence of a hole of radius $d$ around the test particle, 
the integral in (\ref{eq_lambdadef}) may still diverge.
This divergence is an artefact of an approximation employed by Maoz whereby 
stars move along straight lines;
it vanishes if exact Keplerian trajectories are computed for the field stars,
in which case the effective $d$ is roughly $r_G=GM/\sigma^2$ 
\citep{spi87}.
In applying equation (\ref{eq_dfcusp}) to the orbital decay of a single
BH near the center of a stellar system in \S\ref{sec_merge}, 
we take $\ln\Lambda=1/2$ corresponding to
$\gamma=2$ and use for $\rho(d)$ the mean density within a radius
of $0.01$, roughly the radius of gravitational influence of the BH.

The second context in which we applied the dynamical friction formula
in \S\ref{sec_merge} was the orbital decay of two finite-density
spheres representing the original cusps of the merging stellar systems.
We model such a test object with a spherical mass distribution $M(r)$ and 
assume that it is much denser than the field environment.  
According to Gauss' theorem, field particles coming to within pericenter distance $p$ from the center of the test object do not interact with the entire object but only with a portion of mass $M(p)$.  
This suggests an immediate modification of equation (\ref{eq_dfmaoz}):
\begin{equation}
\label{eq_dffinite}
\langle\Delta v_\parallel\rangle=
-\frac{\sqrt{2}G^2}{v\sigma}
\int d^3 r \frac{M(|{\bf R} - {\bf r}|)\rho({\bf r})\alpha}{\left|{\bf R}-{\bf r}\right|^3}
\left[e^{\alpha^2-x^2}\erf(\alpha)-1\right]
\Theta\left(|{\bf R}-{\bf r}|-d\right).
\end{equation}
As an application consider estimating the orbital decay rate for a pair of overlapping Jaffe model galaxies in the final stages of a merger proceeding along a circular orbit.  
Each galaxy has density
\begin{equation}
\rho(r)=\frac{M}{4\pi r_0^3} \left(\frac{r}{r_0}\right)^{-2}
 \left(1+\frac{r}{r_0}\right)^{-2} 
\end{equation}
and thus $M(r)=Mr/r_0$ and $\sigma\approx(GM/2r_0)^{1/2}$.  
Speed of the test body relative to the other body is twice 
the circular velocity $v=2v_c=2(2^{-1/2}\sigma)$, while
$\alpha$ can be expressed terms of the angle $\theta$
between ${\bf r}$ and ${\bf R}$,
\begin{equation}
\alpha=-\frac{xr\sin\theta}{\left|{\bf R}-{\bf r}\right|}.
\end{equation}
Inspection reveals that now, with the test body spread out in space, the limit $d\rightarrow 0$ can be taken inside the integral, hence the $\Theta$-function is identical to unity.

With these substitutions equation (\ref{eq_dffinite}) can be integrated numerically.  We emphasize that treating the test galaxy as a rigid body is a crude approximation; in reality there will be an outer tidal radius beyond which the galaxies are indistinguishable.  The integral depends strongly on the choice of tidal radius.  A natural choice is the separation between the centers $R=a$, which yields
\begin{equation}
\langle\Delta v_\parallel\rangle\approx-\frac{1.50 \sigma^2}{a}.
\end{equation}

\section{Computing Density Profiles}
\label{ap_kernel}

Here we present the algorithms which we used to derive smooth estimates, 
$\hat\nu(r)$ and $\hat\Sigma(R)$, 
of the particle number density and surface density profiles from the 
$N$-body positions.

The routines in {\sf MAPEL\ } \citep{mer94} allow one to derive maximally
unbiased estimates of $\nu$ and $\Sigma$ using penalty functions that embody the approximate power-law nature of these functions.
However the {\sf MAPEL\ } routines are relatively slow,
and this fact presented difficulties when constructing estimates using the $N\sim 10^6$ particle data sets consisting of superposed $N$-body output at several time steps.
Kernel based algorithms are faster but potentially more biased;
however we found them to be adequate for all but the most steeply rising
($\nu\sim r^{-2}$) profiles and so adopted them here.

Our derivation follows that in \citet{met94}.
In the absence of any symmetries in the particle distribution,
a valid estimate of the number density $\nu$ corresponding to 
a set of particle positions ${\bf r}_i$ is
\beq
\hat\nu({\bf r}) = \sum_{i=1}^N {1\over h^3} K\left[{1\over h} \left|{\bf r} - {\bf r}_i\right|\right]
\eeq
where $h$ is the window width and $K$ is a normalized kernel, 
e.g. the Gaussian kernel:
\beq
K(y) = {1\over(2\pi)^{3/2}}e^{-y^2/2}.
\eeq

Now imagine that each particle is smeared uniformly around the surface of the sphere whose radius is $r_i$; typically this sphere will be centered on the single or binary BH.
If the density profile is actually spherically symmetric,
this smearing will leave the density unchanged;
if not, it will produce a spherically symmetric approximation to the true profile.
The spherically-symmetrized density estimate is
\begin{mathletters}
\begin{eqnarray}
\hat\nu(r) &=& \sum_{i=1}^N {1\over h^3} {1\over 4\pi} \int d\phi \int 
d\theta\ \sin\theta\ K\left({d\over h}\right), \\
d^2 &=& \left|{\bf r} - {\bf r}_i\right|^2 \\
&=& r_i^2 + r^2 - 2rr_i\cos\theta 
\end{eqnarray}
\end{mathletters}
where $\theta$ is defined (arbitrarily) from the ${\bf r}_i$-axis.
This may be written in terms of the angle-averaged kernel $\tilde{K}$:
\begin{mathletters}
\begin{eqnarray}
\hat\nu(r) &=& \sum_{i=1}^N {1\over h^3} \tilde{K}(r,r_i,h),\\
\tilde{K}(r,r_i,h) &\equiv& {1\over 4\pi} \int d\phi \int 
d\theta\ \sin\theta\ K\left(h^{-1}\sqrt{r_i^2 + r^2 - 2rr_i\cos\theta}\right)\\
&=& {1\over 2} \int_{-1}^1 d\mu\ K\left(h^{-1}\sqrt{r_i^2 + r^2 - 2rr_i\mu}\right).
\end{eqnarray}
\end{mathletters}
Substituting for the Gaussian kernel, we find
\beq
\tilde{K}(r,r_i,h) = {1\over (2\pi)^{3/2}} \left({r r_i\over h^2}\right)^{-1} e^{-(r_i^2+r^2)/2h^2}\sinh(rr_i/h^2).
\eeq
A better form for numerical computation is
\beq
\tilde{K}(r,r_i,h) = {1\over 2(2\pi)^{3/2}} \left({r r_i\over h^2}\right)^{-1} \left[ 
e^{-(r_i-r)^2/2h^2} - e^{-(r_i+r)^2/2h^2}\right].
\eeq

We want to vary the window width with position in such a way that
the bias-to-variance ratio of the estimate is relatively constant.
Let $h_i$ be the window width associated with the $i$th particle.
The density estimate based on a variable window width is
\beq
\hat\nu(r) = \sum_{i=1}^N {1\over h_i^3} \tilde{K}(r,r_i,h_i).
\eeq
The optimal way to vary $h_i$ is according to Abramson's (1982) rule:
\beq
h_i \propto \nu^{-\alpha}(r_i),\ \ \ \ \alpha = 1/2.
\eeq
Since we don't know $\nu(r_i)$ a priori, 
we compute a pilot estimate of $\nu$ using a fixed kernel
and adjust the $h_i$ based on this estimate \citep{sil86}.

The surface density profile could be computed via simple projection
of $\hat\nu(r)$.
Instead, we computed $\hat\Sigma(R)$ directly from the
coordinates projected along one axis.
The two-dimensional kernel estimate of $\Sigma({\bf R})$ in the
absence of any symmetries is
\beq
\hat\Sigma({\bf R}) = \sum_{i=1}^N {1\over h^2} K'\left[{1\over h} \left|{\bf R} - {\bf R}_i\right|\right]
\eeq
where $K'$ is the two-dimensional Gaussian kernel,
\beq
K'(y) = {1\over 2\pi}e^{-y^2/2}.
\eeq
Now smear each particle uniformly in angle $\phi$ at fixed $R_i$.
The density estimate becomes
\begin{mathletters}
\begin{eqnarray}
\hat\Sigma(R) &=& \sum_{i=1}^N {1\over h^2} {1\over 2\pi} \int  
K'\left({d\over h}\right) d\phi, \\
d^2 &=& R_i^2 + R^2 - 2RR_i\cos\phi . 
\end{eqnarray}
\end{mathletters}
In terms of the angle-averaged kernel $\tilde{K'}$:
\begin{mathletters}
\begin{eqnarray}
\hat\Sigma(R) &=& \sum_{i=1}^N {1\over h^2} \tilde{K'}(R,R_i,h),\\
\tilde{K'}(R,R_i,h) &\equiv& {1\over 2\pi} \int  
 K'\left(h^{-1}\sqrt{R_i^2 + R^2 - 2RR_i\cos\phi}\right) d\phi \\
& = &{1\over 2\pi} e^{-(R_i^2+R^2)/2h^2}I_0(RR_i/h^2)
\end{eqnarray}
\end{mathletters}
where the last expression was derived using the Gaussian kernel;
$I_0$ is the modified Bessel function.

\section{LOSVD Extraction}
\label{ap_losvd}

We carried out non-parametric recovery of the LOSVDs $N(V)$ 
on a dense velocity grid, $-V_0<V<V_0$, 
by maximizing the penalized log-likelihood 
\citep{mer97}
\beq
\log{\mathcal L}_P[N,V_i]=\sum_{i=1}^n \log N(V_i)-\alpha P[N] - 
n\int_{-V_0}^{V_0} N(V) dV
\eeq
where $n$ is the number of particles inside an aperture, 
$V_i$ is the line-of-sight projection of the $i$th particle's velocity,
 and $P[N]$ is a natural choice for the penalty functional that is large 
for noisy $N$ but assigns zero penalty to a Gaussian function \citep{sil86}
\beq
P[N]=\int_{-V_0}^{V_0} \left[\left(\frac{d}{dV}\right)^3\log N(V)\right]^2 dV .
\eeq
When $\alpha$ is very large, maximization of $\log{\mathcal L}_P$ yields a pure Gaussian distribution.  
For the purpose of extracting Gauss-Hermite (GH) moments, 
we chose $\alpha$ smaller than necessary for smooth LOSVDs, 
thereby ensuring that non-Gaussian substructure of the distributions 
is not compromised by smoothing.

Once an LOSVD is available, parameters in the GH expansion can be readily calculated.  
Define the GH moments of $N$ as
\beq
\label{ghmoments}
h_i[N]=2\sqrt{\pi} \int_{-\infty}^{\infty} N(V) g(w) H_i(w) dV
\eeq
where $H_i$ are the Hermite polynomials as defined by \citet{ger93}
and the weight function 
\beq
g(w)=\frac{1}{\sqrt{2\pi}\gamma_0}e^{-w^2/2},\ \ \ \ 
w=\frac{V-V_0}{\sigma_0}
\eeq
has three free parameters ($\gamma_0$, $V_0$, $\sigma_0$).  
Following \citet{mfx93}, 
we choose these parameters such that $h_0[N]=1$ and $h_1[N]=h_2[N]=0$.  
This can be achieved by minimizing the sum $(h_0-1)^2+h_1^2+h_2^2$ as a function of ($\gamma_0$, $V_0$, $\sigma_0$).  
Once these parameters are uniquely determined, 
the higher-order GH moments, including $h_3$ and $h_4$, 
can be evaluated from equation (\ref{ghmoments}) by numerical integration.

\twocolumn

\end{document}